\documentclass[twocolumn]{aastex631}
\hypersetup{linkcolor=red,citecolor=green,filecolor=cyan,urlcolor=magenta}

\newcommand{\cMpch}{$h^{-1}$~cMpc}
\newcommand{\msol}{{\rm M}_{\odot}}
\newcommand{\kms}{km~s${}^{-1}$}

\shorttitle{LATIS: Halo masses from Galaxy-galaxy and Galaxy-Ly$\alpha$ Clustering}
\shortauthors{Newman et al.}

\begin{document}

\title{LATIS: Constraints on the Galaxy-halo Connection at $z\sim 2.5$ from Galaxy-galaxy and Galaxy-Ly$\alpha$ Clustering}

\correspondingauthor{Andrew B. Newman}
\email{anewman@carnegiescience.edu}

\author[0000-0001-7769-8660]{Andrew B. Newman}
\affiliation{Observatories of the Carnegie Institution for Science, 813 Santa Barbara St., Pasadena, CA 91101, USA}

\author{Mahdi Qezlou}
\affiliation{Department of Physics and Astronomy, University of California Riverside, 900 University Ave., Riverside, CA 92521, USA}
\affiliation{Observatories of the Carnegie Institution for Science, 813 Santa Barbara St., Pasadena, CA 91101, USA}

\author{Nima Chartab}
\affiliation{Observatories of the Carnegie Institution for Science, 813 Santa Barbara St., Pasadena, CA 91101, USA}

\author{Gwen C. Rudie}
\affiliation{Observatories of the Carnegie Institution for Science, 813 Santa Barbara St., Pasadena, CA 91101, USA}

\author{Guillermo A. Blanc}
\affiliation{Observatories of the Carnegie Institution for Science, 813 Santa Barbara St., Pasadena, CA 91101, USA}
\affiliation{Departamento de Astronomía, Universidad de Chile, Camino del Observatorio 1515, Las Condes, Santiago, Chile}

\author{Simeon Bird}
\affiliation{Department of Physics and Astronomy, University of California Riverside, 900 University Ave., Riverside, CA 92521, USA}

\author{Andrew J. Benson}
\affiliation{Observatories of the Carnegie Institution for Science, 813 Santa Barbara St., Pasadena, CA 91101, USA}

\author{Daniel D. Kelson}
\affiliation{Observatories of the Carnegie Institution for Science, 813 Santa Barbara St., Pasadena, CA 91101, USA}

\author{Brian C. Lemaux}
\affiliation{Gemini Observatory, NSF’s NOIRLab, 670 N. A’ohoku Place, Hilo,
Hawai’i, 96720, USA}
\affiliation{Department of Physics and Astronomy, University of California, Davis, One Shields Ave., Davis, CA 95616, USA}

\begin{abstract}

The connection between galaxies and dark matter halos is often quantified using the stellar mass-halo mass (SMHM) relation. Optical and near-infrared imaging surveys have led to a broadly consistent picture of the evolving SMHM relation based on measurements of galaxy abundances and angular correlation functions. Spectroscopic surveys at $z \gtrsim 2$ can also constrain the SMHM relation via the galaxy autocorrelation function and through the cross-correlation between galaxies and Ly$\alpha$ absorption measured in transverse sightlines; however, such studies are very few and have produced some unexpected or inconclusive results. We use $\sim$3000 spectra of $z\sim2.5$ galaxies from the Lyman-alpha Tomography IMACS Survey (LATIS) to measure the galaxy-galaxy and galaxy-Ly$\alpha$ correlation functions in four bins of stellar mass spanning $10^{9.2} \lesssim M_* / M_{\odot} \lesssim 10^{10.5}$. Parallel analyses of the MultiDark $N$-body and ASTRID hydrodynamic cosmological simulations allow us to model the correlation functions, estimate covariance matrices, and infer halo masses. We find that results of the two methods are mutually consistent and are broadly in accord with standard SMHM relations. This consistency demonstrates that we are able to accurately measure and model Ly$\alpha$ transmission fluctuations $\delta_F$ in LATIS. We also show that the galaxy-Ly$\alpha$ cross-correlation, a free byproduct of optical spectroscopic galaxy surveys at these redshifts, can constrain halo masses with similar precision to galaxy-galaxy clustering. 

\end{abstract}


\section{Introduction} \label{sec:intro}

Although the luminosity and stellar mass of a galaxy are more readily observed, the mass of its dark matter halo is considered to be the most fundamental parameter for theoretical galaxy evolution models. Measurements of the evolving statistical connection between galaxies' luminosities or stellar masses and their halo masses are therefore crucial for testing and constraining such models. 
 The simplest measure of the galaxy-halo connection is the stellar mass--halo mass (SMHM) relation, which relates the mass of a dark matter halo to its average stellar content.

The main empirical tools for constraining the SMHM relation---and the only ones generally applicable beyond $z \sim 1$---have been measures of galaxy abundances and clustering. The subhalo abundance matching hypothesis \citep{Vale04,Conroy09,Moster10,Guo10,Behroozi10} enables the SMHM relation to be derived from an observed luminosity or stellar mass function by assuming a monotonic relation between these quantities and the halo\footnote{We use ``halo'' to refer to distinct halos and subhalos collectively.} mass or related properties. Galaxy clustering is also used to constrain the SMHM relation by relating the galaxy-galaxy two-point correlation function to that of dark matter halos; more massive halos exhibit stronger clustering \citep[e.g.,][]{Mo96} because they occupy denser environments \citep{Pujol17,Shi18}. The galaxy correlation function can be modeled analytically using a halo occupation distribution (HOD) model \citep{Benson00,Berlind02,Cooray02,Zheng07} or the related conditional luminosity (or stellar mass) function \citep{Yang03,Cooray06,vandenBosch07}. Galaxy clustering may also be modeled by matching directly to subhalos in cosmological simulations \citep{Reddick13,Zheng16,Guo16}. These techniques have been widely applied to galaxy survey data to constrain the galaxy-halo connection \citep[e.g.,][and see references below]{Jing98,Zehavi05,Zheng07,Conroy09,Wake11,Zehavi11,Leauthaud12,Behroozi13,Moster13,Behroozi19,Shuntov22}.

Beyond quasars, studies of galaxy clustering at high redshifts initially focused on Lyman-break galaxies (LBGs; \citealt{Adelberger98,Giavalisco98,Ouchi01,Foucaud03,Adelberger05}), extremely red objects seen in infrared surveys (EROs; \citealt{Daddi00,Moustakas02,Roche02,Zheng04}), and sub-millimeter selected sources (SMGs; \citealt{Webb03,Blain04}). These initial studies showed that LBGs occupy a substantial fraction of halos with masses $M_h \approx 10^{11-12}$~${\rm M}_{\odot}$, whereas the EROs and SMGs occupy somewhat more massive halos. UV-brighter LBGs were found to occupy more massive halos \citep{Giavalisco01,Foucaud03,Ouchi04,Adelberger05,Allen05,Lee06,Hildebrandt09}. With the advent of deep and wide near-infrared imaging surveys, the focus shifted to quantifying stellar mass-dependent clustering at redshifts $z < 2$, with samples usually selected based on photometric redshifts rather than color criteria tailored to galaxy subpopulations \citep{Faucaud10,Wake11,Bielby14,McCracken15}. More recently such efforts have been extended to $z=5$ \citep{Ishikawa17,Cowley18,Shuntov22}. These studies revealed that the SMHM relation retains a similar shape over time, with a peak $M_* / M_h \approx 0.02$ occurring around a characteristic mass $M_h \approx 10^{12} {\rm M}_{\odot}$ that gradually increases with redshift, and they also provided constraints on quenching and the evolution of satellite galaxies.

All of these studies measured the angular correlation function from imaging data in which the distances to most individual galaxies are poorly known. Using a sufficiently large spectroscopic survey, it possible to measure three-dimensional galaxy-galaxy correlation functions. Usually such functions are projected along the line of sight, using well-measured redshifts to eliminate most unassociated galaxy pairs. In addition to measuring the halo mass of the targeted population via its autocorrelation \citep{Bielby13,Trainor12,Durkalec18}, cross-correlations enable the host halo mass of rare objects to be measured (e.g., hyperluminous quasars; \citealt{Trainor12}). Precise measurements of high-redshift galaxy clustering and redshift-space distortions are a premier probe of inflation and early dark energy and motivate massive spectroscopic surveys \citep[e.g.,][]{Schlegel22}.

To date only \citet{Durkalec15,Durkalec18} have measured stellar mass-dependent clustering within a spectroscopic galaxy sample at $z > 2$. Although their initial study \citep{Durkalec15} suggested reasonable agreement with canonical models discussed above, their final analysis based on the full VIMOS Ultra Deep Survey \citep[VUDS,][]{Durkalec18} indicated extremely low halo masses for the lowest-mass galaxies with $M_* \lesssim 10^{9.2}$ ${\rm M}_{\odot}$, suggesting a remarkably high stellar mass fraction $M_* / M_h \approx 0.1$. The authors interpreted this as evidence of very efficient star-formation in low-mass $z \approx 3$ galaxies, which might have been missed by earlier surveys that did reach the lowest masses. If confirmed, this finding would have major implications for galaxy formation models, but it clearly requires verification.

Another tool to constrain the halo masses of high-$z$ galaxies is to measure the cross-correlation between galaxies and the surrounding H~I absorption, as observed in transverse sightlines towards quasars or background galaxies. Excess H~I is found around $z > 2$ galaxies to very large distances. The simplest tracer of the galaxy-H~I cross-correlation is the Ly$\alpha$ flux \citep{Adelberger03,Adelberger05,Crighton11,Bielby17,Chen20}, but Voigt profile fitting \citep{Rudie12} or pixel optical depth methods \citep{Rakic12} can be applied to high-resolution quasar spectra. These studies have examined the galaxy-H~I correlation at small separations to constrain the spatial extent, kinematics, and temperature of inflows and outflows in the circumgalactic region. At separations well beyond $\sim 1$ cMpc (comoving Mpc), on the other hand, the correlation is instead dominated by large-scale structure and is expected to depend mainly on the halo mass \citep{Momose21b}.

\citet{Kim08} made an initial estimate of the halo mass of LBGs based on the surrounding Ly$\alpha$ absorption, although the data of \citet{Adelberger03} permitted only loose constraints with an uncertainty of 0.6 dex. \citet{Rakic13} refined the technique and applied it to an improved data set, deriving a halo mass estimate with a precision of 0.2 dex that was consistent with estimates based on galaxy-galaxy clustering. \citet{Momose21a} was the first to measure the galaxy-Ly$\alpha$ cross-correlation in bins of stellar mass using spectra from the COSMOS Ly$\alpha$ Mapping and Tomography Observations (CLAMATO) survey \citep{Lee18,Horowitz22}. Their models based on a cosmological hydrodynamic simulation showed differences from the measured cross-correlations at distances up to several cMpc, and these were not used to infer halo masses.

Altogether, optical-to-near-infrared imaging surveys have assembled a relatively consistent picture of the SMHM around $z \approx 2$-3, whereas there are very few measurements of stellar mass-dependent galaxy-galaxy or galaxy-Ly$\alpha$ clustering based entirely on spectroscopic surveys at these redshifts, and these have yielded unexpected or inconclusive results. In this paper we use LBG spectra from the Lyman-alpha Tomography IMACS Survey (LATIS; \citealt{Newman20}) to examine both the galaxy-galaxy and galaxy-Ly$\alpha$ correlations consistently. Key advantages of LATIS include a large sample of $\sim$3000 galaxies at $z \approx 2$-3, precise and calibrated redshifts, and coverage of the Ly$\alpha$ forest at a spectral resolving power of $R \sim 900$.

Our goals are (1) to constrain the SMHM relation at $z = 2.5$ using galaxy-galaxy and galaxy-Ly$\alpha$ clustering, both to inform the galaxy-halo connection generally and as a key input to future papers examining the galaxy-IGM connection using LATIS, (2) to assess the consistency of the methods as a precise test of the methodology and of systematic errors in the LATIS Ly$\alpha$ forest data, and (3) to evaluate the relative precision of galaxy-galaxy and galaxy-Ly$\alpha$ clustering as constraints on halo occupation to help inform future surveys. 

We introduce our galaxy-galaxy clustering analysis in Section 2. We then turn to measurements of the galaxy-Ly$\alpha$ cross-correlation (Section~\ref{sec:gallyaobs}), calculation of the halo-Ly$\alpha$ cross-correlation in a cosmological simulation (Section~\ref{sec:gallyasim}), and the resulting constraints on halo masses (Section~\ref{sec:gallyaresults}). We compare the SMHM relations from galaxy-galaxy and galaxy-Ly$\alpha$ clustering to one another and to the literature (Section~\ref{sec:discussion}) and summarize their implications (Section~\ref{sec:summary}).

\section{Galaxy-Galaxy Clustering}
\label{sec:galgalclustering}

We begin with our methods for estimating the observed galaxy and simulated (sub)halo correlation functions. A key challenge is the construction of an accurate covariance matrix, which we address using mock surveys within a large cosmological simulation. After developing the method, we first estimate the typical halo mass of the LATIS galaxies, undifferentiated by stellar mass, based on their autocorrelation function. This measurement is required for some analyses in which the galaxy sample is not divided by mass, e.g., \citet{Newman22}. Second, we consider subsets of galaxies defined by stellar mass and derive halo masses using a new method that incorporates all $N(N+1)/2$ auto- and cross-correlations among the $N$ subsamples.

\subsection{Observations}
\label{sec:galgalobsdata}

\begin{figure}
    \centering
    \includegraphics[width=3.5in]{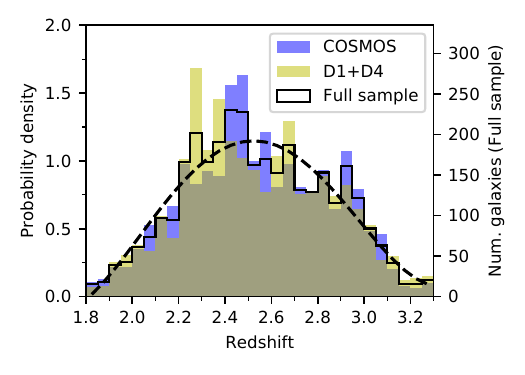}\vspace{-2ex}
    \caption{Redshift distribution of galaxies used to measure galaxy-galaxy clustering. The full sample is shown along with the COSMOS and D1+D4 subsamples plotted separately to illustrate field-to-field variation. Each histogram is a normalized probability density function; the right axis shows the corresponding number of galaxies per bin in the full sample. The dashed line shows the polynomial fit used to generate random points.\label{fig:zpdf}}
\end{figure}

We measure the clustering of LATIS LBGs with high-confidence ({\tt zqual} $=3$ or 4) redshifts $z=2$-3, approximately the central 90\% of the redshift distribution shown in Fig.~\ref{fig:zpdf}, combining all three survey fields (COSMOS, CFHTLS-D1, and CFHTLS-D4). Redshifts are derived from full spectral modeling, and for the high-confidence subset used in this paper, we visually identified multiple spectral features. \citet{Newman20} estimated that the catastrophic error rate for the high-confidence redshifts was $<2\%$, which we will consider as negligible. We exclude galaxies that were observed only as part of ``bright target'' masks intended for poor weather conditions \citep{Newman20}, since they do not cover the entire survey area; galaxies whose spectra are heavily blended; quasars; and targets not covered by the near-infrared imaging required to robustly estimate stellar masses, which affects only the CFHTLS-D4 field.

\begin{figure*}
    \centering
    \includegraphics[width=6in]{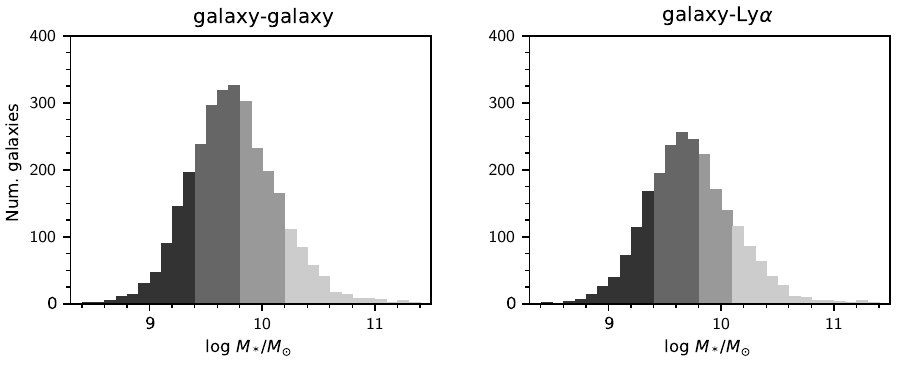}
    \caption{Stellar mass distributions of galaxies used in the galaxy-galaxy (left panel) and galaxy-Ly$\alpha$ (right) clustering analyses. Different shadings indicate the four stellar mass bins.}
    \label{fig:massdist}
\end{figure*}

The photometric catalogs and derivation of stellar masses from spectral energy distribution fitting are described by Chartab et al (2023, in press). Fig.~\ref{fig:massdist} shows the stellar mass distribution. Although its full range spans $\log M_* / {\rm M}_{\odot} = 8.4$-11.4, galaxies are concentrated around the median $M_* = 10^{9.7}$ ${\rm M}_{\odot}$, and the central 90\% range is $\log M_* / {\rm M}_{\odot} = 9.1$-10.4, a factor of 20. The total number of galaxies used to measure galaxy-galaxy clustering is 2991.

\subsection{Clustering Estimator}

The two-point correlation function $\xi(r_p, \pi)$ is the excess probability of finding a galaxy at a transverse separation $r_p$ and line-of-sight separation $\pi$ (in redshift space) from another galaxy, relative to a uniformly distributed population. We will primarily use the two-point correlation function projected along the line of sight:
\begin{equation}
w_p(r_p) = \int_{-\pi_{\rm max}}^{\pi_{\rm max}} \xi(r_p, \pi) d\pi.
\end{equation}
Note that in this convention $w_p$ has units of distance. The integral limits $\pm\pi_{\rm max}$ are generally selected to be large enough to desensitize the measurement to redshift-space distortions and galaxy redshift errors, while being small enough to avoid adding excessive noise from uncorrelated pairs. Typical values of $\pi_{\rm max}$ are 15-30 $h^{-1}$ cMpc \citep[e.g.,][]{Trainor12,Marulli13,Kashino17,Durkalec18}, and we set $\pi_{\rm max} = 20$ $h^{-1}$ cMpc. Our mock surveys, described below, showed that the precision of the inferred halo masses is not very sensitive to the choice of $\pi_{\rm max}$ within this range.

To estimate $\xi(r_p, \pi)$, we generate a large number of unclustered random points (see below), convert the celestial coordinates and redshifts of the observed galaxies and random points into comoving Cartesian coordinates, and use the {\tt Corrfunc} code \citep{Sinha20} to rapidly compute the \citet{Landy93} estimator. In the case of an autocorrelation, the simple version of this estimator is
\begin{equation}
\xi(r_p, \pi) = \frac{DD - 2DR + RR}{RR},
\end{equation}
where $DD$, $DR$ and $RR$ are the normalized number of galaxy-galaxy, galaxy-random, and random-random, respectively, within a given bin of $r_p$ and $\pi$. These counts are normalized to account for the oversampling of random points: $DD = N_{DD} / [N_D (N_D - 1)]$, $DR = N_{DR} / (N_D N_R)$, and $RR = N_{RR} / [N_R (N_R - 1)]$, where $N_{DD}$, $N_{DR}$, and $N_{RR}$ represent the raw pair counts \citep[e.g.,][]{Kerscher00}. In the case of a cross-correlation between two galaxy or halo samples, we have
\begin{equation}
\xi(r_p, \pi) = \frac{D_1 D_2 - D_1 R_2 - D_2 R_1 + R_1 R_2}{R_1 R_2},
\end{equation}
where the normalized counts are $D_1D_2 = N_{D_1 D_2} / (N_{D_1} N_{D_2})$, $D_1R_2 = N_{D_1 R_2} / (N_{D_1} N_{R_2})$ (analogous for $D_2R_1$), and $R_1 R_2 = N_{R_1 R_2} / (N_{R_1} N_{R_2})$, and $N_{D_1 D_2}$, $N_{D_1R_2}$, $N_{D_2R_1}$, and $N_{R_1R_2}$ are the raw galaxy-galaxy, galaxy-random, and random-random pair counts. Once the two-dimensional correlation function $\xi(r_p, \pi)$ has been estimated, we integrate along the line of sight to obtain $w_p(r_p)$. Numerically the integration consists of summing estimates of $\xi$ over $\pi$ bins with a width of 1 $h^{-1}$ cMpc.

An essential aspect of estimating $\xi$ is the generation of random coordinates of unclustered objects that must respect the survey geometry and redshift distribution. For each of the three survey fields, we randomly draw $N_{R} = 100 \times N_{D}$ celestial coordinates within the survey area, excluding masked regions around bright stars (see Appendix~\ref{sec:esr}), where $N_D$ is the number of observed galaxies. The redshift distribution of an unclustered population would not be uniform due to the LATIS target selection function, which depends on a galaxy's $r$-band magnitude and colors. The best estimate of the selection function is empirical: the redshift pdf of the observed sample. Although this estimate inevitably and undesirably includes large-scale structure, its influence can be mitigated by averaging over independent fields and smoothing the distribution \citep[e.g.,][]{Trainor12,Marulli13}. Fig.~\ref{fig:zpdf} shows the redshift pdf constructed in bins of $\Delta z = 0.05$ and modeled with a polynomial of degree $n = 4$, which we use to draw random samples. We find that our results are extremely insensitive to $n$ for any reasonable value $n \geq 2$. Although only $z=2$-3 galaxies are used to estimate $\xi$, we include galaxies over a slightly wider redshift range when fitting this polynomial to better constrain the endpoints.

We use a weighting scheme to account for non-uniformity in the survey sensitivity and sampling. Appendix~\ref{sec:esr} describes our estimates of the target sampling rate (TSR) and spectroscopic success rate (SSR) as a function of celestial coordinates. The TSR gives the fraction of the LATIS parent sample that was observed, while the SSR gives the fraction of the observed sample for which a high-confidence spectroscopic redshift was determined. The product of these rates is the effective sampling rate, or ESR. The ESR exhibits a variety of angular dependencies arising from slitmask design constraints, instrumental effects, and time-dependent observing conditions. Similar to other studies \citep[e.g.,][]{Marulli13} we account for angular variation of the ESR by weighting galaxy $i$ by $w_i = {\rm ESR}^{-1}$. Specifically, in the weighted estimator of $\xi(r_p, \pi)$, the pair counts above are replaced by sums of products of weights $\sum {\rm ESR}^{-1}_i \times {\rm ESR}^{-1}_j$, and all random points receive the mean weight. (Recall that random points are already constrained to lie within the observed survey area.) 

Beyond contributing to these large-angle variations in the ESR, slitmask design constraints lead to a deficit of close galaxy pairs in LATIS relative to the parent sample. However, this deficit requires no correction because it exceeds 5\% only for separations less than 0.3~arcmin (see also \citealt{Newman22}), which is smaller than any separation bin that we will consider.

These procedures are initially used to estimate $w_p(r_p)$ in each of the three LATIS fields separately, and we then compute a final $w_p(r_p)$ from a weighted mean over the fields, where field $i$ receives a weight $\mathcal{W}_i$. In the limit of Poisson errors, inverse variance weighting would assign a weight proportional to the number of pairs in a given radial bin. In the limit that $r_p$ is much smaller than the survey volume, this expected pair number is proportional to $n_i^2 V_i$, where $n_i$ is the galaxy space density and $V_i$ is the volume of field $i$. We therefore use $\mathcal{W}_i = n_i^2 V_i / \sum_{i=1}^3 n_i^2 V_i$; the weights for the COSMOS, D1, and D4 fields are 0.59, 0.36, and 0.05, respectively. Experiments with the mock surveys introduced below shows that these simple weights are close to optimal in terms of minimizing variance.

\subsection{Halo Correlation Functions}

We compute halo correlation functions using the MultiDark Planck 2 simulation (MDPL2; \citealt{Klypin16}), which was selected to enable a sufficient number of mock surveys (see below) distributed within its 1 $h^{-3}$ {\rm Gpc}${}^3$ volume. We use the snapshot at $z_{\rm sim}=2.535$, which is very close to the mean redshift of the galaxies $\langle z \rangle = 2.52$. We compute the correlation functions of all halos (i.e., distinct halos and subhalos) in the {\tt RockStar} \citep{Rockstar} catalog, as appropriate for our comparison to galaxy clustering; for reference, about 8\% of halos with masses $M_h = 10^{11.7} M_{\odot}$ are subhalos. We define $M_h$ using the virial mass {\tt Mvir}. Due to the small expected fraction of subhalos in the relevant mass and redshift range along with our limited constraining power on sub-Mpc scales, we prefer our single-parameter approach over fitting a more flexible HOD model, which could suffer from significant degeneracies.

First, we compute $w_p(r_p)$ for thresholded samples of halos, i.e., those with a virial mass $M_{\rm vir} > M_h^{\rm thresh}$ for a selected $M_h^{\rm thresh}$. We find the positions of these halos in redshift space by perturbing their comoving $z$ coordinates by $v_z (1+z) / H(z)$, where $v_z$ is the physical halo velocity in the $z$ direction. For speed, we randomly select a subsample of at most $5 \times 10^5$ halos, which we find is sufficient to estimate $w_p$ to $\sim 1$\% precision. We use {\tt Corrfunc} to compute $w_p$ on a grid of $\log M_h^{\rm thresh} / \msol$ spanning 10.7 to 13.5 in steps of 0.1, and we use linear interpolation to evaluate $w_p$ between grid points. 

Second, we consider the cross-correlations between two halo samples, each consisting of an essentially ``mono-mass'' halo population defined by a narrow range $\pm0.1$~dex around a specified mass $M_h$. We compute $w_p(r_p)$ on a two-dimensional grid and use bilinear interpolation where necessary.

Our galaxy sample spans a range of redshifts $z=2$-3 over which clustering evolves, whereas the halo clustering is calculated at a single redshift. In the limit of linear growth and biasing, the halo autocorrelation function is
\begin{equation} 
\xi_{\rm h}(z) = b(z)^2 \xi_{\rm dm}(z) 
= \left[\frac{b^2(z) D^2(z)}{b^2(z_{\rm sim}) D^2(z_{\rm sim})}\right] \xi_{\rm h}(z_{\rm sim}),
\end{equation}
where $b(z)$ is the bias, $D(z)$ is the linear growth factor, and $\xi_{\rm dm}$ is the dark matter correlation function. We find that $\langle b^2(z_i) D^2(z_i) \rangle / [b^2(z_{\rm sim}) D^2(z_{\rm sim})]$ = 1.0002, where the average is taken over the redshifts $z_i$ of galaxies in our sample, which means that we can directly compare the galaxy and halo autocorrelation functions when the full galaxy sample is considered. However, when we consider auto- and cross-correlations of galaxies in bins of stellar mass, we make small corrections for clustering evolution, because the mean redshift of the galaxies varies slightly with stellar mass. We do this by rescaling the halo $w_p$ by a factor $f = b(z_1) D(z_1) b(z_2) D(z_2) / [b^2(z_{\rm sim}) D^2(z_{\rm sim})]$, where $z_1$ and $z_2$ are the mean redshifts of the galaxy subsamples. We use the \citet{Tinker10} bias model and the {\tt Colossus} code \citep{Diemer18} to evaluate $f$, which is quite weakly sensitive to halo mass. We find that this amounts to a very minor correction with $0.976 < f < 1.019$.

\subsection{The Integral Constraint}

Consider the average value of an estimator $\xi^{\rm est}$ of the correlation function evaluated over random pairs of points within a survey volume. Because $\xi^{\rm est}$ is an excess probability relative to random pairs, this average must be zero regardless of the true value $\langle \xi^{\rm true} \rangle$ that would obtain over an infinite volume. The difference is known as the integral constraint or IC \citep{Infante94,Roche02}. The IC is approximately a negative additive bias that can be estimated by averaging $\xi^{\rm true}$ over random pairs in the survey volume; the projected correlation function then satisfies $w^{\rm est}_p(R) \approx w^{\rm true}_p(R) - {\rm IC} \times 2\pi_{\rm max}$. We estimate the IC as part of our fitting process by evaluating it for each proposed halo population \citep{Wake11}. Although the IC is often a significant correction to angular correlation functions, it has far less impact on a spectroscopic survey, because the distance between random points is dominated by the survey's line-of-sight depth, which is much larger than $\pi_{\rm max}$. We correct for the IC but find that it changes $w_p$ by at most a few percent in the outermost radial bins.

\begin{figure*}
\centering
\includegraphics[width=7in]{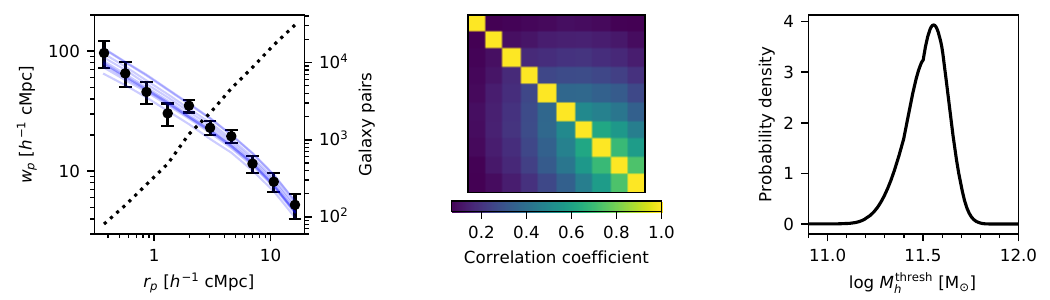}
\caption{\emph{Left:} The projected galaxy-galaxy correlation function $w_p(r_p)$ for the entire LATIS sample, undivided by stellar mass. Black points show measured values with error bars $C_{ii}^{0.5}$ derived from the mock surveys. These data are modeled by the autocorrelation function of mass-thresholded halos; blue curves show 10 posterior samples. The dotted line and right axis show the number of galaxy pairs (within $|\Delta z| < \pi_{\rm max}$) in each $r_p$ bin. \emph{Middle:} The correlation matrix $C_{ij} / (C_{ii} C_{jj})^{0.5}$ derived from mock surveys. Each row or column corresponds to a bin of $r_p$, increasing down and to the right. \emph{Right:} Posterior probability density for the threshold halo mass $M^{\rm thresh}_h$. \label{fig:autocorr}}
\end{figure*}

\subsection{Clustering of the Full Galaxy Sample}
\label{sec:galgalwholesample}

Figure~\ref{fig:autocorr} (left panel) shows the autocorrelation function of the LATIS galaxies, computed in 10 logarithmic bins of $r_p$ spanning 0.3-20 $h^{-1}$ cMpc. Our first aim is to model the galaxy clustering with a thresholded halo sample defined by $M_h > M^{\rm thresh}_h$. The main difficulty in such modeling is that $w_p$ errors in different radial bins are correlated, and on sufficiently large scales, this correlation is dominated by large-scale structure. As in many prior clustering studies \citep[e.g.,][]{delaTorre13,Durkalec15}, we use mock surveys to estimate the data covariance matrix. These mock surveys require an initial estimate of  $M^{\rm thresh}_h$, which we derive by assigning independent Poisson errors to the $w_p$ measurement in each radial bin, and then fitting these data with our set of halo $w_p(r_p)$ profiles using a maximum likelihood procedure. The initial estimate is $\log M^{\rm thresh}_{h,0} / \msol = 11.56$.

For each LATIS survey field, we then select non-overlapping subvolumes that fill the MDPL2 simulation box. For the largest field, there are 135 such subvolumes. With each subvolume, we assign a mock redshift $z_0$ to each halo such that the comoving line-of-sight distance from $z_{\rm min} = 2$ to $z_0$ matches the distance of the halo to the front of the simulation box. We also assign mock celestial coordinates $({\rm RA}_0, {\rm Dec}_0)$ to each halo by converting its $x$ and $y$ coordinates to angular transverse distances, appropriate to the halo's mock redshift, from a sightline passing through the center of the subvolume; the celestial coordinates of this sightline are set to the center of the observed field. These calculations assume the distant observer approximation. We then randomly select halos with masses $M_h > M^{\rm thresh}_{h,0}$ with a probability proportional to ${\rm ESR}({\rm RA}_0, {\rm Dec}_0) \times z_{\rm pdf}(z_0)$. The total number of selected halos matches the total number of LATIS galaxies in the field. We then use the same code to compute $w_p$ that is applied to the observed data. This process is repeated to realize 100 random halo selections per subvolume. Finally, we select many random triplets $(i, j, k)$, where $i$, $j$, and $k$ identify a realization of a subvolume associated with each of the three survey fields, and compute the mean $w_p$ using the same weights applied to the observations. From these realizations we estimate the covariance matrix $C_{ij}$ of $w_p$. The correlation matrix $C_{ij} / (C_{ii} C_{jj})^{1/2}$ is shown in Fig.~\ref{fig:autocorr}. Note that the bins are increasingly correlated at larger $r_p$.

We can now perform a Bayesian inference on $M^{\rm thresh}_h$. We take a broad, uniform prior on $\log M_h$ and assume a Gaussian likelihood with the covariance matrix $C_{ij}$ estimated from the mock surveys. The resulting posterior is shown in Fig.~\ref{fig:autocorr} (right panel). We summarize the pdf using the mode and the 16th and 84th percentiles: $\log M^{\rm thresh}_h / \msol = 11.56^{+0.06}_{-0.15}$. The mock surveys showed the pdf mode to be an unbiased estimator, whereas the median is biased low because the pdfs are left skewed. The minimum $\chi^2 = 8.3$ for 9 degrees of freedom, so deviations from the model are consistent with the uncertainties. Since our inferred $M^{\rm thresh}_h$ matches the initial estimate $M^{\rm thresh}_{h,0}$ (but with much more realistic errors), we did not find it necessary to iterate this procedure. 

To facilitate comparisons with future work, we also fit a power law model, though we stress that this plays no role in our halo mass inference. If the three-dimensional correlation function follows a power law $\xi(r) = (r / r_0)^{-\gamma}$, then the projected correlation function $w_p$ can be expressed as a power law multiplied by a hypergeometric function, which appears due to the finite integration window over $\pi$ (see \citealt{Trainor12}, Equation 10, which must be multiplied by $2 \pi_{\rm max}$ to match our $w_p$ convention).  We fix $\gamma = 1.5$ since, like \citet{Trainor12}, we found this value to match the shape of the halo autocorrelation function within the relevant mass range, and measure $r_0 = 4.9 \pm 0.3$ \cMpch.

\subsection{Stellar Mass-dependent Clustering}

\begin{figure*}
    \centering
    \includegraphics[width=0.8\linewidth]{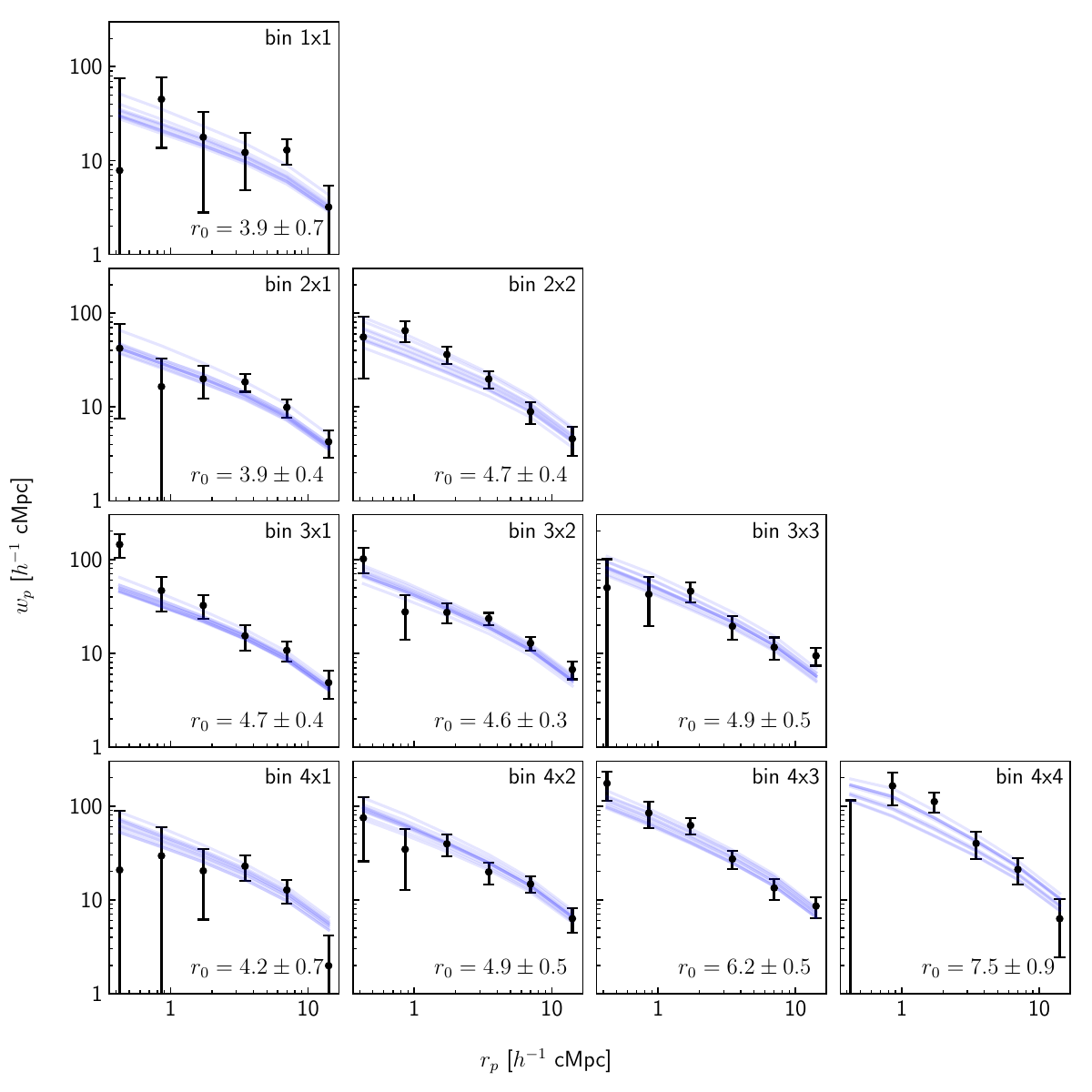}
    \caption{The projected auto- and cross-correlations among LATIS galaxies in four bins of stellar mass listed in Table~\ref{tab:galgalmasses}. Each panel shows the measured $w_p(r_p)$ (black points) with $C_{ii}^{0.5}$ error bars based on the covariance matrix $C_{ij}$ derived from mock surveys. Blue curves show 10 posterior samples of the halo model. The correlation length $r_0$, expressed in \cMpch, is shown in each subpanel.}
    \label{fig:crosscorr}
\end{figure*}

We now consider the clustering of LATIS galaxies as a function of their stellar mass. A common method to analyze mass-dependent galaxy clustering is to compute $w_p$ for a series of stellar mass-thresholded subsamples, i.e., $M_* > M^{\rm thresh}_*$, and to fit such data with an HOD model \citep[e.g.,][]{Wake11,Durkalec18}. This approach has drawbacks for analyzing high-redshift spectroscopic samples. First, no such sample is stellar mass-limited; in the case of LATIS, the stellar mass distribution has a shape closer to Gaussian than a truncated mass function. Therefore observed galaxies at relatively undersampled masses need to be up-weighted, and this introduces a dependence on the assumed stellar mass function. Second, in the high stellar mass bins, the threshold approach incorporates only the pairs of massive galaxies with themselves, omitting their many pairs with lower-mass galaxies. This is not optimal when sample sizes are limited.

We introduce a different method in which galaxies are divided into $N_b$ bins of stellar mass, and we compute and simultaneously model all of the auto- and cross-correlations (hereafter simply cross-correlations) between the $N_b \times (N_b+1) / 2$ pairs of bins. This approach incorporates all galaxy pairs into the analysis. Figure~\ref{fig:crosscorr} shows the 10 cross-correlations arising from the 4 stellar mass bins listed in Table~\ref{tab:galgalmasses}. Given the smaller number of galaxies in the individual mass bins, we calculate $w_p(r_p)$ using a reduced set of 6 $r_p$ bins.

\begin{deluxetable}{cccccc}
    \tablewidth{\linewidth}
    \tablecolumns{6}
    \tablecaption{Halo mass constraints from galaxy-galaxy clustering}
    \startdata
    \tablehead{\colhead{Bin num.} & \colhead{$\log M_*$} & \colhead{$\langle \log M_* \rangle$} & \colhead{$N_{\rm gal}$} & \colhead{$\langle z \rangle$} & \colhead{$\log M_h$}}
    1 & $< 9.4$ & 9.21 & 547 & 2.58 & $11.39^{+0.19}_{-0.36}$ \\
    2 & 9.4-9.8 & 9.61 & 1181 & 2.53 & $11.78^{+0.11}_{-0.18}$ \\
    3 & 9.8-10.2 & 9.97 & 899 & 2.49 & $12.03^{+0.12}_{-0.12}$ \\ 
    4 & $> 10.2$ & 10.46 & 364 & 2.47 & $12.44^{+0.13}_{-0.22}$
    \enddata
    \tablecomments{Masses are expressed in $\msol$.\label{tab:galgalmasses}}
\end{deluxetable}

We compare these galaxy cross-correlations to halo cross-correlations, modeling each galaxy stellar mass bin as a mono-mass halo population. The procedure is analogous to that described in Section~\ref{sec:galgalwholesample}. We first estimate an initial $M_{h,0}^i$ for each of the 4 bins, indexed by $i$, by using Poisson error estimates and simultaneously fitting the 10 cross-correlations. We then build mock surveys; in each realization, we now select 4 random samples of halos around $M_{h,0}^i$, with the number of halos matching the number of observed galaxies in stellar mass bin $i$. Halos are selected according to the $z_{\rm pdf}$ described by galaxies in the associated stellar mass bin. The covariance matrix $C_{ij}$ is calculated as before, but it is now $60 \times 60$. We then proceed to a Bayesian inference, as before. To sample the posterior, we use {\tt emcee} \citep{emcee}. The main difference compare to Section~\ref{sec:galgalwholesample} is that we find that three iterations are necessary: after each iteration, we use the resulting estimates of the halo masses to create a new suite of mock surveys. After three iterations, the inferred halo masses are sufficiently converged ($\Delta \log M_h \leq \sigma/4$, where $\sigma$ is the random error). The change in $\log M_h$ between the first and last iterations is $\leq 0.08$ in all stellar mass bins; therefore, by iteratively bringing the mocks and observations into agreement, we improve internal consistency but ultimately make a small refinement to the halo masses.

Figure~\ref{fig:crosscorr} shows that all cross-correlations can be simultaneously modeled within their uncertainties, with a minimum $\chi^2 = 50.1$ for 56 degrees of freedom. The corner plot in Appendix~\ref{sec:corners} shows that the posteriors are unimodal and covariance among the 4 halo mass parameters is low. The halo masses that we infer (Table~\ref{tab:galgalmasses}) increase monotonically with stellar mass. These results are consistent with our analysis of the LATIS sample as a whole (Section~\ref{sec:galgalwholesample}): above a threshold of $M_h^{\rm thresh} = 11.56$, the median halo mass is $\log M_h = 11.80$, which is consistent with the interpolation of the halo mass estimates in Table~\ref{tab:galgalmasses} to the median $\log M_* = 9.72$.  In Section~\ref{sec:discussion}, we will return to these measurements and compare them to the literature and our Ly$\alpha$ analysis.

To facilitate comparisons with future work and to help clarify the significance of differences among the stellar mass bins in Fig.~\ref{fig:crosscorr}, we also fit a power law model (see Section~\ref{sec:galgalwholesample}) to each of the the auto- and cross-correlation functions independently. Values of $r_0$ assuming a fixed $\gamma = 1.5$ are shown in each subpanel.

\section{Galaxy-Ly$\alpha$ Cross-correlation: Observations}
\label{sec:gallyaobs}

We now turn to the second method for measuring halo masses: the galaxy-Ly$\alpha$ cross-correlation. In this section we describe the observations and measurements, while Section~\ref{sec:gallyasim} describes our calculation of the halo-Ly$\alpha$ cross-correlation from a cosmological hydrodynamic simulation, and Section~\ref{sec:gallyaresults} presents the inference procedure and the resulting halo mass estimates.

\begin{figure*}
\centering
\includegraphics[width=3in]{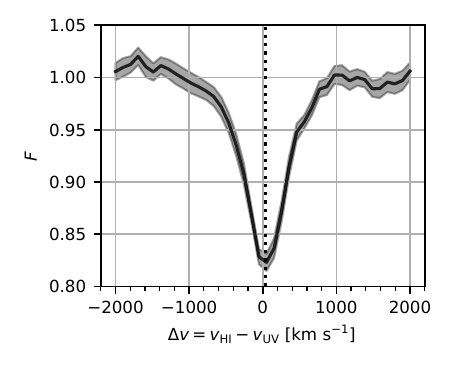}
\includegraphics[width=3in]{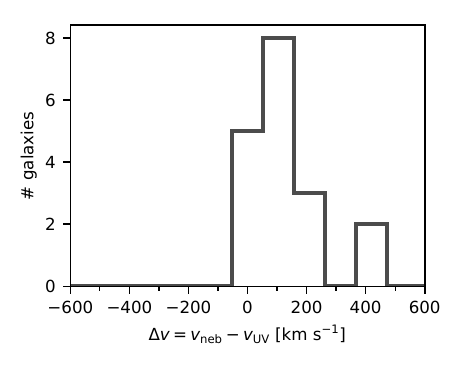}
\vspace{-0.5cm}
\caption{Calibration of LATIS redshifts using transverse absorption and nebular emission. \emph{Left:} Average spectrum (and 1$\sigma$ uncertainty band) of sightlines within $r_p < 4$ $h^{-1}$ cMpc of LATIS galaxies, shifted into each galaxy's rest frame using the LATIS UV redshifts. We detect a small asymmetry around $\Delta v = 0$ attributed to a systematic offset in the UV redshifts. \emph{Right:} Comparison of LATIS UV redshifts to MOSDEF nebular redshifts for 18 galaxies in common.}
\label{fig:zabs}
\end{figure*}

\subsection{Ly$\alpha$ Forest Spectra}
\label{sec:sightlines}

We measure Ly$\alpha$ absorption in a set of 3007 sightlines toward galaxies and quasars. Among the full LATIS sample, we select objects with $z > 2.277$ so that the Ly$\alpha$ forest is at least partly observed, and we exclude sources identified as having major data reduction flaws, poor fits of the spectral models that describe the unabsorbed continuum, and severe blending.  The processing of spectra was described by \citet{Newman20}. The key measurement is the Ly$\alpha$ transmission fluctuation
\begin{equation}
\delta_F = \frac{F}{\langle F \rangle(z)} - 1,
\end{equation}
where $F$ is the flux at a given wavelength normalized by the unabsorbed continuum model (see \citealt{Newman20}), and $\langle F \rangle(z)$ is the redshift-dependent mean flux transmitted through the IGM \citep{FG08}. We consider only spectral pixels for which the uncertainty $\sigma_{\delta_F} < 2$. This leaves $4.7 \times 10^5$ Ly$\alpha$ forest pixels in our data set.

There are two further processing steps. First, we attempt to exclude high-column-density absorption lines from our cross-correlation. Although not an essential step, we find that their inclusion does not substantially improve constraining power, but it increases model dependence because subgrid models are required to simulate these lines accurately. We use a simple technique that masks lines with a high equivalent width (EW; see \citealt{Newman20}). Briefly, we convolve the continuum-normalized spectrum with 5 pixel boxcar, identify local minima, and measure the EW. The EW is measured within the smallest wavelength interval over which $F < \langle F \rangle(z)$, truncated to a maximum of $\pm 1000$ km s${}^{-1}$ from the minimum in order to avoid identifying broad, shallow features. If the rest-frame ${\rm EW} > 5$~\AA~and the feature is significantly detected (${\rm EW} / \sigma_{\rm EW} > 5$), the line is masked over the wavelength interval where $F < \langle F \rangle(z)$. Based on the simulations described in Section~\ref{sec:gallyasim}, we find that this simple method identifies 95\% of lines with column densities $N_{\rm HI} > 10^{20.7}$ cm${}^{-2}$; it inevitably also masks some lower-$N_{\rm HI}$ lines that are blended at our spectral resolution. 

Second, we refine the unabsorbed continuum model for each sightline using a multiplicative polynomial that forces $\langle \delta_F \rangle \approx 0$ on large scales. The polynomial order depends on the length of the Ly$\alpha$ forest that is covered by the spectrum (see \citealt{Newman20}), but the filtering scale is typically $\Delta z \approx 0.2$. This process is called mean flux regularization (MFR, \citealt{Lee12}). MFR mitigates errors in the spectrophotometric calibration, but it also suppresses large-scale correlations.

It is important to bear in mind that we measure a processed version of $\delta_F$, and that the same procedures that exclude high-EW lines and perform MFR will be applied to the simulated Ly$\alpha$ forest spectra to enable an accurate comparison.

\subsection{Foreground Galaxies and Redshift Estimates}
\label{sec:foregr} 

The foreground sample consists of the galaxies whose positions will be cross-correlated with Ly$\alpha$ absorption. We use a subset of the galaxy sample described in Section~\ref{sec:galgalobsdata}, restricting to the redshift range $z_{\rm fg} = 2.22$-2.90. The lower limit ensures that the region with $\pm 2000$ km s${}^{-1}$ of Ly$\alpha$ lies within our observed spectral range, while the upper limit is motivated by the rapidly declining number of background sources.

Rest-UV redshifts rely mainly on the resonant Ly$\alpha$ line (in emission and absorption) and metallic absorption lines produced in part by outflowing gas; neither traces the galaxy systemic redshift directly. Rest-UV surveys often have systematic velocity offsets compared to indicators of galaxies' systemic redshifts. Although these are irrelevant to galaxy-galaxy clustering, such offsets would affect the galaxy-Ly$\alpha$ cross-correlation. One way to estimate a systematic velocity offset is to examine the average transverse Ly$\alpha$ absorption profile, which should be symmetric about galaxies' systemic redshifts \citep{Rakic12}. For each sightline-foreground galaxy pair with a transverse separation $r_p < 4$ $h^{-1}$ cMpc, we shift the continuum-normalized spectrum into the galaxy rest frame using the LATIS redshift (see \citealt{Newman20}), and we then compute the inverse-variance weighted mean over pairs (Figure~\ref{fig:zabs}). We find that the absorption is nearly symmetric about $\Delta v = v_{\rm HI} - v_{\rm UV} = 0$, but we do detect a small offset $\Delta v = 39 \pm 16$~\kms. 
~The offset shows no clear trend with spectral morphology as indicated by the Ly$\alpha$ EW, so we simply add the same $\Delta v$ to all LATIS redshifts.

We can also evaluate systematic offsets, as well as random errors, in the LATIS redshifts by comparing them to nebular redshifts for 18 galaxies in common with the MOSDEF survey \citep{Kriek15}. We find that the median offset of $\Delta v = v_{\rm neb} - v_{\rm UV} = 96 \pm 36$~\kms~is consistent with, but less precise than, the estimate from transverse Ly$\alpha$ absorption. (The uncertainties are determined from bootstrap resampling.) The standard deviation of velocity difference is $\sigma_v = 120 \pm 24$~\kms, which may be taken as an error in the UV redshifts given that the errors in MOSDEF redshifts are expected to be much smaller; however, inspecting the histogram in Figure~\ref{fig:zabs} shows that $\sigma_v$ is heavily influenced by the two largest outliers. The LATIS redshifts for these two galaxies are very sensitive to whether the Ly$\alpha$ region is included in the spectral model fit; if it is not, the redshifts obtained from the interstellar absorption lines agrees well with the MOSDEF redshifts. This situation is rare: a comparably large ($> 300$~\kms) dependence on the inclusion of Ly$\alpha$ occurs in only 0.7\% of the LATIS sample, which indicates that the subset in common with MOSDEF is likely unrepresentative. Excluding those two galaxies, the estimated redshift error becomes $\sigma_v = 66 \pm 8$~\kms. Such an exclusion is reasonable but still uncomfortably ad hoc. We use these results only to estimate a reasonable prior on $\sigma_v$ for our galaxy-Ly$\alpha$ analysis: we will use a normal distribution centered on the average of our two estimates (120 and 66 \kms) with a dispersion set to their half-difference, i.e., $93 \pm 27$~\kms.

\subsection{Estimating the Galaxy-Ly$\alpha$ Cross-Correlation}
\label{sec:gallyaestimating}

To estimate the cross-correlation between LATIS galaxies and Ly$\alpha$ absorption, we follow a similar procedure as \citet{FontRibera12}, who measured the cross-correlation between damped Ly$\alpha$ absorbers and Ly$\alpha$ fluctuations measured in quasar spectra. The cross-correlation is defined as
\begin{equation}
\xi^{\rm Ly \alpha}(r_p, \pi) = \frac{\sum_{i \in A} w_i \delta_{F,i}}{\sum_{i \in A} w_i},
\end{equation}
where the summations are over a list $A$ of those pixels located within a separation bin $(r_p \pm \Delta r_p, \pi \pm \Delta \pi)$ from a galaxy in the foreground sample; note the same pixel may appear multiple times. This is simply a weighted mean of $\delta_F$, and $\xi^{\rm Ly\alpha}$ can be equivalently thought of as a Ly$\alpha$ absorption profile. We use inverse variance weights that incorporate both observational noise and the intrinsic variance in the Ly$\alpha$ forest:
\begin{equation}
w_i = \left[ \sigma_F^2(z_i) + \sigma_{{\rm rand},i}^2 + \sigma_{{\rm cont},i}^2 \right]^{-1},
\label{eqn:weights}
\end{equation}
where $\sigma_F^2(z_i)$ is the intrinsic variance at the redshift $z_i$ of pixel $i$, $\sigma_{{\rm rand},i}$ is the random noise in $\delta_{F,i}$, and $\sigma_{{\rm cont},i}$ is the estimated noise in $\delta_{F,i}$ attributed to errors in the continuum model. We approximate $\sigma_F(z) = 0.20 \left[(1+z) / 3.5\right]^{2.4}$ based on the simulated spectra that we introduce below, and therefore $\sigma_F^2$ represents the total variance in noiseless spectra with the same spectral resolution and processing as the LATIS data. We evaluate $\sigma_{{\rm cont}, i}$ as a function of the continuum-to-noise ratio following \citet[][Section 7.4]{Newman20}. Note that these weights naturally prevent the dominance of a few spectra with high signal-to-noise ratio.

\begin{figure*}
    \includegraphics[width=7in]{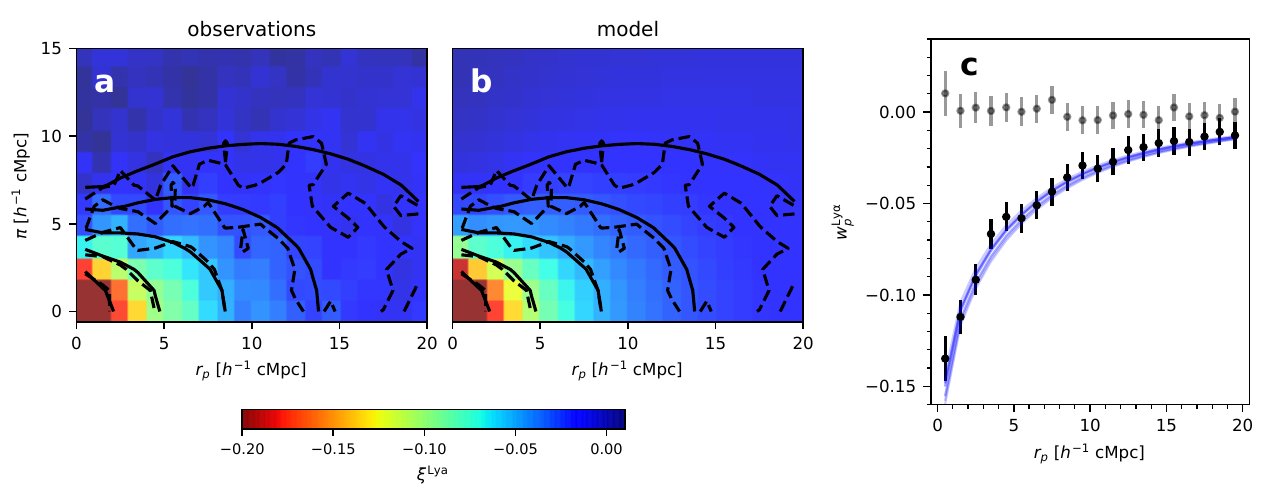}
    \caption{The galaxy-Ly$\alpha$ cross-correlation $\xi^{\rm Ly\alpha}$ computed from the full LATIS foreground galaxy sample, independent of stellar mass. \emph{Panel a:} The observed 2D correlation function. Contours begin at $\xi^{\rm Ly\alpha} = -0.2$ and decrease to -0.0125 by factors of 2. Dashed and solid contours indicate the observed and model correlation functions, respectively. \emph{Panel b:} The model 2D correlation function; contours are the same as in panel a. \emph{Panel c:} The projected cross-correlation $w_p^{\rm Ly\alpha}$. Black points show observed values with $1\sigma$ errors derived from the covariance matrix, and blue curves show 10 posterior samples of the model. Gray points show a null test constructed by scrambling the Ly$\alpha$ forest spectra among the sightlines. \label{fig:gallya_onebin_model}}
\end{figure*}

We use linear bins of $r_p$ spanning 0-20 $h^{-1}$ cMpc with widths of 1 $h^{-1}$ cMpc and bins of $\pi$ spanning -15 to 15 $h^{-1}$ cMpc with widths of 1.2 $h^{-1}$ cMpc, which approximately matches the LATIS pixel scale (1.8~\AA) at the mean redshift. We then fold the cross-correlation about $\pi = 0$, averaging estimates in the fore- and background; therefore the first $\pi$ bin covers 0 to 0.6 $h^{-1}$ cMpc. 

Fig.~\ref{fig:gallya_onebin_model}a shows the 2D galaxy-Ly$\alpha$ cross-correlation derived from the entire LATIS sample, independent of stellar mass. A strong detection of excess Ly$\alpha$ absorption to large distance is clearly apparent, along with an anisotropic shape. We will discuss the 2D structure of the cross-correlation further in Section~\ref{sec:gallyasim}.

It is useful to consider a projected cross-correlation function akin to the $w_p$ statistic that we employed to measure galaxy-galaxy clustering. We define $w_p^{\rm Ly\alpha}$ by \emph{averaging} $\xi^{\rm Ly\alpha}$ within $|\pi| < \pi_{\rm max} = 6.6$ $h^{-1}$ cMpc:
\begin{equation}
w_p^{\rm Ly\alpha}(r_p) = \frac{1}{2\pi_{\rm max}} \int_{-\pi_{\rm max}}^{\pi_{\rm max}} \xi^{\rm Ly\alpha}(r_p, \pi) d\pi.
\end{equation}
Numerically this corresponds to averaging the first 6 $\pi$ bins (see Fig.~\ref{fig:gallya_onebin_model}) with half the weight given to the first bin as to the others, since it is centered on 0. Note that this convention differs from that for the galaxy-galaxy $w_p$, which is an integral over $\pi$ rather than an average. The disadvantage of our definition for $w_p^{\rm Ly\alpha}$ is that it does not converge as $\pi_{\rm max} \to \infty$, but we find it convenient for $w_p^{\rm Ly\alpha}$ to maintain a dimensionless form to facilitate comparisons with systematic uncertainties in $\delta_F$. No single choice of $\pi_{\rm max}$ is optimal at all $r_p$, but based on the simulations introduced below, we find that the signal-to-noise ratio of $w_p^{\rm Ly\alpha}$ is close to maximal over a wide range of $r_p$ for our choice of $\pi_{\rm max}$.

When analyzing galaxy-galaxy clustering we had to account for angular variations in the ESR arising from any source, since they directly affect the number of pairs that are counted. In LATIS such variations mainly depend on a galaxy's position within the IMACS field of view. Such sampling variations are not relevant to the galaxy-Ly$\alpha$ cross-correlation, since they do not bias the measured $\delta_F$. Potentially concerning are variations in ESR that correlate with local density, and thus $\delta_F$. \citet{Newman22} simulated recovery of galaxy density in a two-pass LATIS-like survey, respecting geometrical constraints imposed by slitmask design. On 2-8 $h^{-1}$ cMpc scales, where $\xi^{\rm Ly\alpha}$ is best constrained, the galaxy density was biased $\lesssim 2\%$ within density fluctuations $< 3\sigma$. Only a small fraction of LBGs reside in higher-density regions, so we can safely assume that any such sampling bias has a negligible effect on $\xi^{\rm Ly\alpha}$.

\subsection{Data Covariance Matrix}
\label{sec:gallyacov}

Measurements of $\xi^{\rm Ly\alpha}$ in different bins are highly covariant for several reasons. First, a given Ly$\alpha$ forest pixel forms pairs with many galaxies and so appears in many bins. Second, continuum errors are coherent along sightlines. Third, Ly$\alpha$ forest fluctuations are correlated by large-scale structure. We follow \citet{FontRibera12} to estimate the covariance matrix analytically.

Consider two separation bins $A$ and $B$ of $\xi^{\rm Ly\alpha}$. The covariance of the cross-correlation $\tilde{C}_{AB}$ is
\begin{equation}
    \tilde{C}_{AB} = \frac{\sum_{i \in A} \sum_{j \in B} w_i w_j C_{ij}}{\sum_{i \in A} w_i \sum_{j \in B} w_j}
\end{equation}
(Eqn.~3.11 of \citealt{FontRibera12}) where $C_{ij} = \langle \delta_{F,i} \delta_{F,j} \rangle $ is the correlation of $\delta_F$ measured in pixels $i$ and $j$. Taking the transverse and perpendicular separations of these pixels to be $r_{p, ij}$ and $\pi_{ij}$, we estimate the correlation as
\begin{equation}
    C_{ij} = \xi_F(r_{p,ij}, \pi_{ij}) + \sigma^2_{\rm rand, i} \delta_{ij}^K + \sigma^2_{\rm cont, i} \delta^D(r_{p,ij})
\end{equation}
(analogous to Eqn.~3.12 of \citealt{FontRibera12}). The first term is the intrinsic autocorrelation of $\delta_F$, which we compute directly from the simulations introduced below. This term encodes cosmic variance. The second term represents the random noise in the spectra, which appears only with $i = j$ such that the Kronecker delta function $\delta^K_{ij} = 1$ ($\delta^K_{ij} = 0$ when $i \neq j$). The last term represents the continuum error, which appears only when the pixels are members of the same sightline so that the Dirac function $\delta^D(r_{p,ij}) = 1$ ($\delta^D(r_{p,ij}) = 0$ when $r_{p,ij} \neq 0$). We approximate the continuum error as being perfectly correlated along a sightline with a variance $\sigma^2_{\rm cont}$ that is a function of the median continuum-to-noise ratio along the sightline (see Section 7.4, \citealt{Newman20}). Thus $\sigma^2_{{\rm cont}, i} = \sigma^2_{{\rm cont}, j}$ whenever the third term is active. Given the covariance matrix $\tilde{C}$ of $\xi^{\rm Ly\alpha}$, it is straightforward to calculate the covariance matrix of the projected $w_p^{\rm Ly\alpha}$.

The matrix $\tilde{C}$ is slow to compute. Each bin of $\xi^{\rm Ly\alpha}$ is formed by averaging pixels from many galaxy-Ly$\alpha$ pairs. Computing $\tilde{C}_{AB}$ then requires calculating a weighted mean of $C_{ij}$ over every pair of pixels in bins $A$ and $B$. To speed the calculation, we parallelize it and approximate $\xi_F = 0$ when $r_{p,ij}$ or $|\pi_{ij}|$ exceeds 30 $h^{-1}$ cMpc.

\section{Galaxy-Ly$\alpha$ Cross-correlation: Simulations}
\label{sec:gallyasim}

In order to estimate a halo mass from the observed galaxy-Ly$\alpha$ cross-correlation, we compute simulated halo-Ly$\alpha$ cross-correlations using the large and high-resolution ($2 \times 5500^3$ particles in a volume of $250^3$ $h^{-3}$ cMpc${}^3$) cosmological hydrodynamic simulation ASTRID \citep{Bird22,Ni22}. In this section we describe our construction of Ly$\alpha$ skewers from ASTRID, the processing of these skewers into mock LATIS spectra, and the calculation of redshift- and mass-dependent halo-Ly$\alpha$ cross-correlations.

\subsection{Ly$\alpha$ Optical Depths}
Our construction of Ly$\alpha$ absorption spectra from the ASTRID simulations follows the procedures of \citet[][Section 2.2 and 3.1]{Qezlou22, Qezlou23}. Briefly, each gas particle in the simulation is treated as an individual absorber with internal physical properties smoothed by a quintic spline kernel. The absorption spectrum is then composed of all the Voigt profiles of the absorbers along a given line of sight. We estimate the neutral hydrogen fraction by assuming a uniform ultraviolet background (UVB) and solving the collisional/photoionization and recombination rate networks from \citet{Katz96}. The self-shielding of dense neutral gas is modeled using fitting formulae derived from radiative transfer models \citep{Rahmati13}. We effectively set the UVB intensity by scaling the Ly$\alpha$ optical depths to match the mean flux evolution observed by \citet{FG08}, as corrected for metal absorption. We compute a grid of $1000^2$ sightlines parallel to the $z$ axis with a transverse spacing of 0.25 \cMpch. The spectral pixel size is also 0.25 \cMpch, or about 27 \kms. This procedure is repeated for snapshots at $z = 2.3$, 2.5, 2.7, and 2.9, approximately matching the range of our foreground galaxy sample.

\subsection{Halo Masses in ASTRID versus MultiDark}
\label{sec:hmdef}

\begin{figure}
    \centering
    \includegraphics[width=3in]{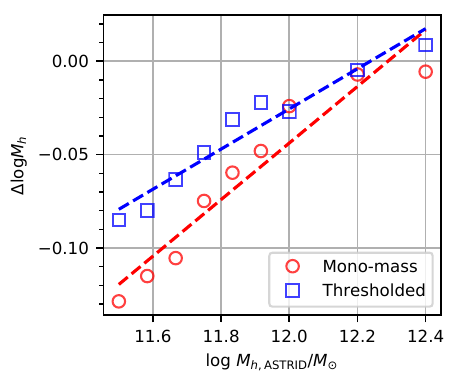}
    \caption{Differences in the masses of halo populations with matched autocorrelation functions in MDPL2 and ASTRID. For each halo population in ASTRID (mono-mass: $|\log M_h /  M_{h,\rm ASTRID}| < 0.1$, thresholded: $M_h > M_{h, {\rm ASTRID}}$), we adjust the mass parameter $M_{h, \rm MDPL2}$ in MDPL2 to achieve the best match to $w_p$, defined as the rms fractional difference over $1 < r_p / (h^{-1} {\rm cMpc})  < 10$ $h^{-1}$. Linear fits are shown: $\Delta \log M_h = \log M_{h, \rm MDPL2} / M_{h, \rm ASTRID} = a(\log M_{h, \rm ASTRID} / 10^{12} {\rm M}_{\odot}) + b$, with $a, b = 0.15, -0.04$ (mono-mass) or $0.11, -0.03$ (thresholded).\label{fig:hmdef}}
\end{figure}

Among our goals is to compare the consistency of halo masses estimated from galaxy-galaxy and galaxy-Ly$\alpha$ clustering. We use different simulations that are most appropriate for each analysis: the large, $N$-body simulation MDPL2 to enable a large suite of mocks for galaxy-galaxy clustering, and the smaller but hydrodynamic simulation ASTRID to compute Ly$\alpha$ spectra. These simulations are based on slightly different cosmological parameters, particularly $\sigma_8$, and different methods were used to construct the subhalo catalogs: {\tt RockStar} \citep{Rockstar} for MDPL2 and {\tt SUBFIND} \citep[][see \citealt{Bird22}]{Springel01} for ASTRID.

We present halo masses derived from each method (galaxy-galaxy and galaxy-Ly$\alpha$ clustering) based on the associated simulation. However, when we compare the results from these methods in Section~\ref{sec:discussion}, we will account for the expected mass difference arising from differences in the simulations. We estimate this mass difference by comparing halo-halo clustering in MDPL2 and ASTRID. Fig.~\ref{fig:hmdef} shows that halo populations with matching autocorrelation functions $w_p$ have slightly different masses in MDPL2 and ASTRID, with $|\Delta \log M_h| < 0.1$ at masses relevant for our analysis. In other words, if we were able to self-consistently conduct both analyses within ASTRID, we expect the halo masses from galaxy-galaxy clustering would be slightly higher, by $|\Delta \log M_h|$, than those we inferred using MDPL2.

\subsection{Synthesizing Mock LATIS Spectra}

The high-resolution synthetic spectra are then processed to mimic the LATIS data. First, we smooth the spectra by the LATIS line spread function. Based on the average extent of galaxies measured in the two-dimensional spectra, the line spread function can be approximated as a gaussian with $\sigma / ({\rm km~s}^{-1}) = 227 - 98 (\lambda / {\rm 500~nm})$. We evaluate this expression for each snapshot at the observed wavelength of Ly$\alpha$. The convolved spectra are then resampled to a velocity grid with a cell size that  matches the LATIS pixel size of 1.8~\AA~in the observed frame.

The spectra must then be further processed to account for the masking of high-EW lines and MFR (Section~\ref{sec:sightlines}). Our ultimate aim of extracting the mean Ly$\alpha$ absorption signal around halos does not directly require us to inject noise. However, correctly capturing the average effect of the high-EW line masking procedure does require the spectra to have noise properties similar to LATIS. Therefore we temporarily inject random noise and continuum errors into the synthetic spectra. The random noise is gaussian white noise with a dispersion randomly chosen for each sightline by drawing a pixel from LATIS at a redshift close to that of the snapshot. We also inject continuum error by multiplying $F$ by a sinusoid scaled so that its variance is $\sigma^2_{\rm cont}$, which in turn is a function of the sightline's continuum-to-noise ratio \citep{Newman20}. We find that our results are not sensitive to the details of the continuum error modeling; they have only an indirect effect of modulating which high-EW lines are masked, and these are relatively rare and do not dominate the mean absorption signal.

We then identify the high-EW lines and perform MFR on the noisy synthetic spectra, following methods applied to the observations (Section~\ref{sec:sightlines}). One complication is that the MFR polynomial order depends on the observed forest length, which is a distribution that varies with redshift, whereas the synthetic sightlines all span the full box. We mimic MFR as follows. For a given snapshot and sightline, we first draw a random LATIS background galaxy whose Ly$\alpha$ forest probes the snapshot redshift, and we compute the observed length $L$ of the forest as limited by our blue cutoff $\lambda > 3890$~\AA. Starting from a random position along the sightline, we partition it into (1 or 2) chunks of length $L$, wrapping around the periodic boundary when necessary, and then perform MFR separately on each. In many cases $L$ exceeds the ASTRID box size, and the simulation is therefore missing power on scales that are observed. However, MFR removes power on scales $\Delta z \lesssim 0.3$, depending on $L$, while the depth of the ASTRID box is $\Delta z = 0.31$ at $z=2.5$. Therefore the missing power would be removed by MFR anyway, and we conclude that the simulation box is large enough to adequately mimic MFR. In a small fraction $7 \times 10^{-5}$ of sightlines, MFR makes a large $> 4\times$ correction to the continuum level that can result in extreme values of $\delta_F$ and even unphysically negative continuua. We mask these rare sightlines; they have no counterparts in the observed data set.

Finally we apply these high-EW line masks and MFR polynomials to the noiseless synthetic spectra. We are left with $10^6$ sightlines per snapshot that mimic the LATIS resolution, sampling, and processing.

\subsection{Halo-Ly$\alpha$ Cross-correlations}
\label{sec:profiles}

\begin{figure*}
    \centering
    \includegraphics[width=0.8\linewidth]{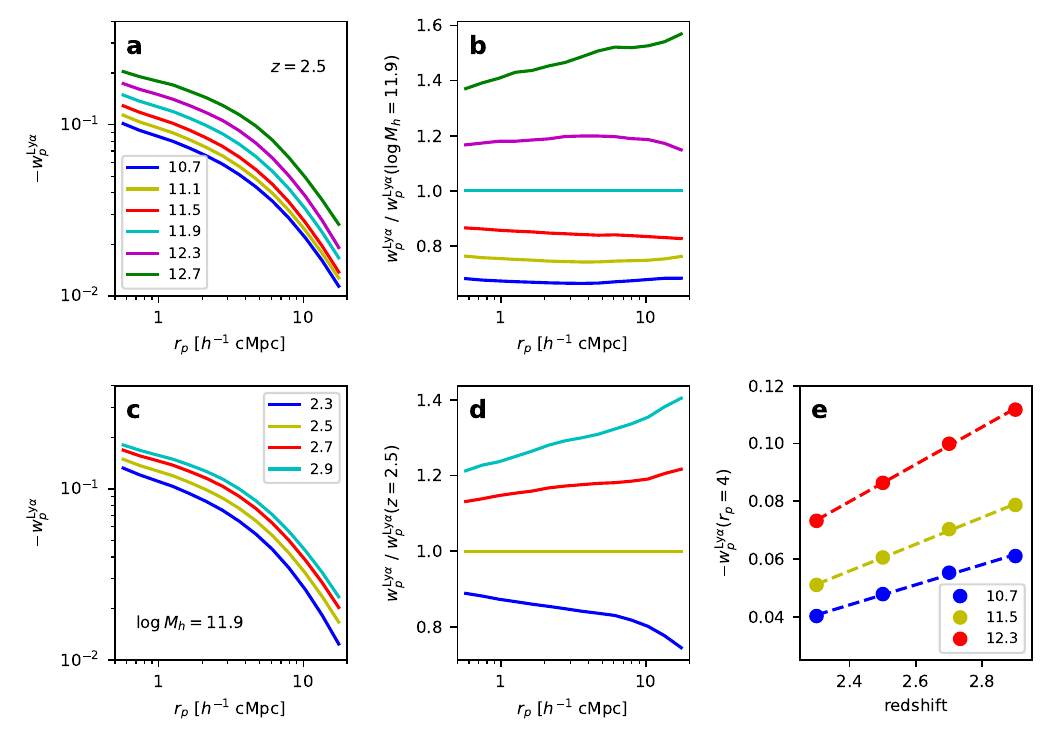}
    \vspace{-3ex}
    \caption{Trends in the projected halo-Ly$\alpha$ cross-correlation with halo mass and redshift. \emph{Panel a:} Variation in $w_p^{\rm Ly\alpha}(r_p)$ with halo mass ($\log M_h / {\rm M}_{\odot}$ indicated in legend) in the $z = 2.5$ simulation snapshot. \emph{Panel b:} The ratio of each curve in panel a to the $M_h = 10^{11.9} {\rm M}_{\odot}$ curve, highlighting the weak dependence of the shape of $w_p^{\rm Ly\alpha}(r_p)$ on $M_h$. \emph{Panel c:} Variation of $w_p^{\rm Ly\alpha}(r_p)$ with redshift at the fixed halo mass indicated. \emph{Panel d:} Ratio of each curve in panel c to the one at $z = 2.5$. \emph{Panel e:} Variation of $w_p^{\rm Ly\alpha}$ with redshift at fixed $M_h$ (indicated in legend) and fixed $r_p = 4$ \cMpch, highlighting the linearity of the redshift dependence. \label{fig:model_trends}}
\end{figure*}

We next compute the average Ly$\alpha$ absorption around halos as a function of their mass and redshift. As in our galaxy-galaxy clustering analysis, we consider both thresholded samples with $M_h > M^{\rm thresh}_h$ and mono-mass samples with masses in a narrow range $|\log(M_h / M_{h,0})| < 0.1$.  
For a given snapshot and halo sample, we first randomly select a subsample of up to $5 \times 10^5$ halos. We compute the redshift-space position of each halo and round it to the nearest point defined by the grid of synthetic sightlines. (In the spectral direction, this nearest-neighbor assignment mimics our procedure of assigning LATIS pixels to the nearest $\pi$ bin when constructing $\xi^{\rm Ly\alpha}$.) We extract a 3D cube of $\delta_F$ measurements around this point, and we average these cubes over the halo subsample. We verified that the absorption is centered in the resulting flux cube. We then project each flux cube into a 2D ($r_p, \pi$) grid with the same cell sizes as the observed $\xi^{\rm Ly\alpha}$, averaging over $r_p$ and using spline interpolation over $\pi$.

Fig.~\ref{fig:model_trends} illustrates the main trends. We first consider $w_p^{\rm Ly\alpha}$ for ease of visualization. Panels a and b show that, at a fixed redshift, Ly$\alpha$ absorption around halos is approximately scale-independent, i.e., it differs by a factor that depends on $M_h$ but little on $r_p$. Only at the highest masses $M_h \gtrsim 10^{12.7} M_{\odot}$ does the shape of $w_p(r_p)$ appreciably change. Panels c and d show that at fixed $M_h$, the shape of $w_p(r_p)$ varies with redshift. When we fix both $M_h$ and $r_p$ and examine the dependence on redshift (panel e), we find that it is quite linear over the range $z=2.3$-2.9. This permits a simplification: we need not consider the distribution of redshifts of our foreground galaxy sample when fitting a model; only the mean redshift is relevant.

\begin{figure}
    \centering
    \includegraphics[width=3.0in]{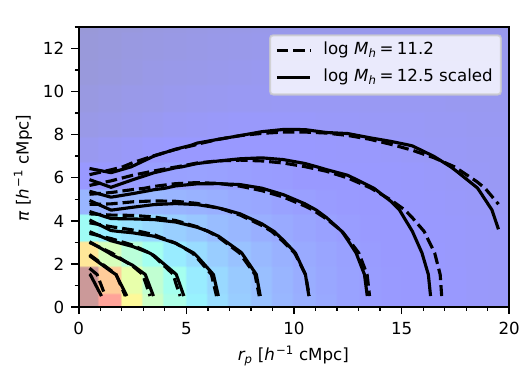}
    \vspace{-1ex}
    \caption{The insensitivity of the shape of the halo-Ly$\alpha$ correlation function $\xi^{\rm Ly\alpha}$ to halo mass is demonstrated by comparing $\xi^{\rm Ly\alpha}(M_h = 10^{11.2} \msol)$ (colors following Fig.~\ref{fig:gallya_onebin_model}, and dashed contours) to $0.54 \times \xi^{\rm Ly\alpha}(M_h = 10^{12.5} \msol)$ (solid contours). Both are computed at $z=2.5$.\label{fig:xi_2D_model}}
\end{figure}

The spatial distribution of Ly$\alpha$ absorption is insensitive to the halo mass not only in projection, but also in its two-dimensional structure, as noted by \citet{Kim08}. Figure~\ref{fig:xi_2D_model} shows that contours of $\xi^{\rm Ly\alpha}(M_h)$, at fixed redshift, have an almost identical shape even for halo masses that differ by a factor of 20. It is only the amount of Ly$\alpha$ absorption that constrains the halo mass, not its radial or line-of-sight distribution.

\section{Galaxy-Ly$\alpha$ Cross-Correlation: Fitting and Results}
\label{sec:gallyaresults}

Finally we compare our observed 2D and 1D galaxy-Ly$\alpha$ cross-correlations, $\xi^{\rm Ly\alpha}$ and $w_p^{\rm Ly\alpha}$, to the computed halo-Ly$\alpha$ cross-correlations and infer halo masses. Fitting $\xi^{\rm Ly\alpha}$ in 2D should provide the most constraining power, not because the redshift-space anisotropies it encodes are sensitive to halo mass (see Section~\ref{sec:profiles}), but simply because it is analogous to a matched filter. Our simple definition of $w_p^{\rm Ly\alpha}$, which uses fixed integration limits $\pi_{\rm max}$, cannot provide an optimal estimator of the total absorption at every $r_p$ and so must lose some statistical power. On the other hand, the projected cross-correlation function $w_p^{\rm Ly\alpha}$ has the merits of being insensitive to the modeling of redshift errors and redshift-space distortions; it should be less precise but possibly more robust. We will fit both and compare results. As in our galaxy-galaxy clustering analysis, we model the full LATIS galaxy sample with a thresholded halo population, and galaxy subsamples binned by stellar mass with mono-mass halo populations.

\subsection{Inference}

Consider a division of the LATIS foreground galaxies into $N$ stellar mass bins: we will use $N=1$ for the full sample and $N=4$ for the stellar mass-defined subsamples. We begin by describing the inference of halo masses from the 2D cross-correlation $\xi^{\rm Ly\alpha}$. Since $\xi^{\rm Ly\alpha}$ has 20 bins of $r_p$ and 13 bins of $\pi$, the length of the data vector is 260$N$, and the covariance matrix $\tilde{C}$ described in Section~\ref{sec:gallyacov} is $260N \times 260N$. There are $N+1$ parameters: $N$ describing the halo mass $M_h$ associated with each galaxy bin, and one additional parameter $\sigma_z$ that encodes random errors in the galaxy redshifts, which is assumed to be the same for all mass bins. For each mass bin $i$ and estimated $\xi^{\rm Ly\alpha}_i$, we compute the weighted mean redshift $\langle z \rangle_i$ of all the $\delta_F$ measurements that entered $\xi^{\rm Ly\alpha}_i$ using the same weights (Equation~\ref{eqn:weights}). As justified by Figure~\ref{fig:model_trends}e, we linearly interpolate between snapshots to produce a grid of synthetic $\xi^{\rm Ly\alpha}$ that is matched to $\langle z \rangle_i$ and varies over $M_h$. We then linearly interpolate over this grid to the target $\log M_{h,i}$. Finally we convolve the resulting model along the $\pi$ direction using a gaussian kernel with dispersion $\sigma_z$. We compute a standard gaussian likelihood $\mathcal{L}$ from the difference between the data and model vectors and the covariance matrix $\tilde{C}$. We take broad uniform priors on $\log M_{h, i} / M_{\odot}$ spanning 10.7-12.9 and the gaussian prior on $\sigma_z$ described in Section~\ref{sec:foregr}. The posteriors are then sampled using {\tt emcee}. 

We also infer halo masses from the projected cross-correlation function $w^{\rm Ly\alpha}_p$. Motivated by maximizing the robustness of this analysis, we further eliminate the innermost bin with $r_p < 1$ $h^{-1}$ cMpc where feedback is expected to have the strongest impact. This reduces the data vector size to $19N$; it is straightforward to compute its associated covariance matrix from $\tilde{C}$. 

\subsection{Results: Full Sample}

Fig.~\ref{fig:gallya_onebin_model} shows the results of fitting $\xi^{\rm Ly\alpha}$ as measured on the full foreground galaxy sample, undifferentiated by mass, to a thresholded halo sample. Visually the model provides an excellent fit both to the observed 2D structure (compare panels a and b) and the projected absorption profile (panel c). Some systematic differences are apparent beyond $r_p \sim 10$ \cMpch, but these measurements are highly correlated and, as we will show, the differences are not significant. We find $\log M^{\rm thresh}_h / M_{\odot} = 11.5 4 \pm 0.09$ and $\sigma_z = 65 \pm 17$~\kms~(see corner plot in Appendix~\ref{sec:corners}). Note that the $\sigma_z$ posterior is narrower than and shifted from the center of the prior, indicating that the galaxy-Ly$\alpha$ clustering favors redshift errors at the lower end of our range of estimates based on comparing to nebular redshifts (Section~\ref{sec:foregr}). As a metric of fit quality, we note that at the MAP parameters, $\chi^2 = 255$ for a data vector of size 260, indicating that the errors are well described by our covariance matrix.

We also performed a fit to the projected $w_p^{\rm Ly\alpha}$ alone. The resulting halo mass estimate $\log M^{\rm thresh}_h / M_{\odot} = 11.49 \pm 0.13$ is slightly lower but still consistent with that obtained from the two-dimensional $\xi^{\rm Ly\alpha}$. The uncertainty in $\log M^{\rm thresh}_h$ increases in this case, as expected, by about 50\%, and the posterior of $\sigma_z$ matches the prior, since all sensitivity to redshift errors was destroyed by the projection.

\begin{deluxetable*}{cccccc}
    \tablewidth{\linewidth}
    \tablecolumns{6}
    \tablecaption{Halo mass constraints from galaxy-Ly$\alpha$ clustering}
    \startdata
    \tablehead{\colhead{Bin number} & \colhead{$\log M_*$ range} & \colhead{$\langle \log M_* \rangle$} & \colhead{$\langle z \rangle$} & \colhead{$\log M_h$} (2D fit) & \colhead{$\log M_h$ (1D fit)}}
    1 & $< 9.4$ & 9.21 & 2.51 & $11.75^{+0.17}_{-0.14}$ & $11.82^{+0.16}_{-0.26}$ \\
    2 & 9.4-9.8 & 9.61 & 2.48 & $11.86^{+0.08}_{-0.11}$ & $11.74^{+0.12}_{-0.20}$ \\
    3 & 9.8-10.1 & 9.93 & 2.45 & $11.86^{+0.09}_{-0.17}$ & $12.02^{+0.11}_{-0.22}$\\ 
    4 & $> 10.1$ & 10.36 & 2.45 & $11.80^{+0.12}_{-0.17}$ & $12.07^{+0.09}_{-0.30}$
    \enddata
    \tablecomments{Masses are expressed in $\msol$. 2D and 1D fits are to $\xi^{\rm Ly\alpha}$ and $w_p^{\rm Ly\alpha}$, respectively.\label{tab:gallyamasses}}
\end{deluxetable*}

\begin{figure*}
    \centering
   \includegraphics[width=\linewidth]{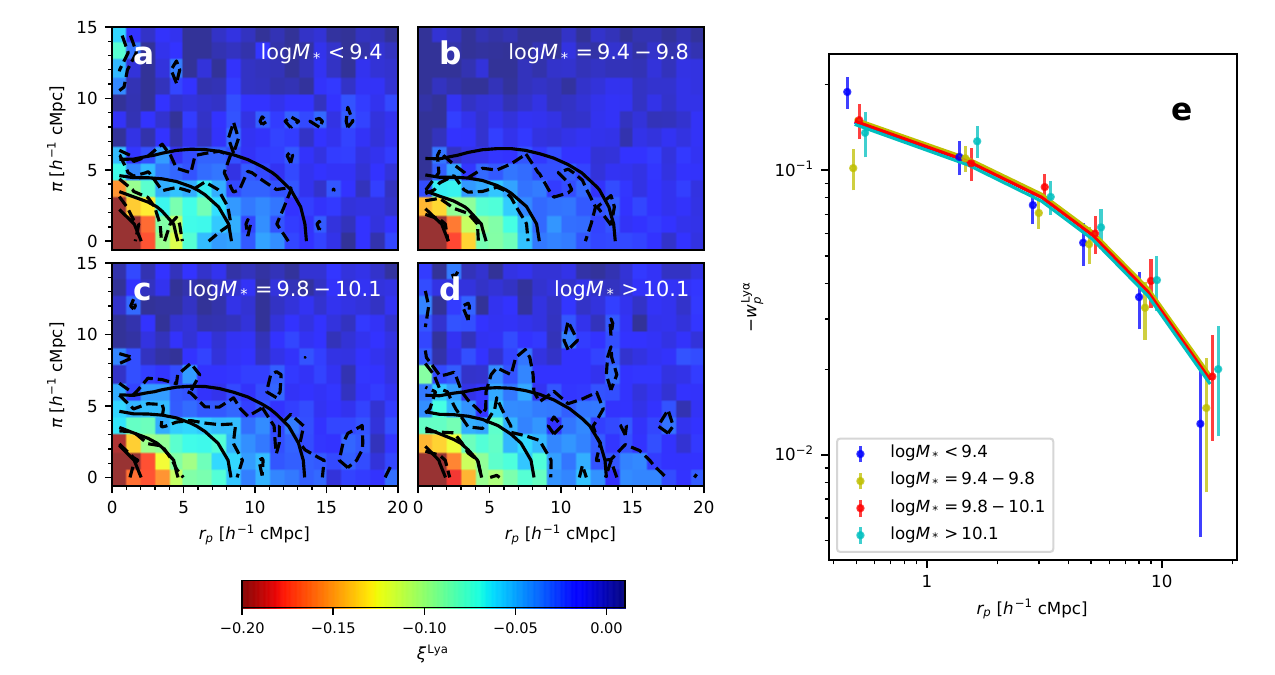}
    \caption{\emph{Panels a-d:} The galaxy-Ly$\alpha$ correlation function measured in four bins of $M_*$. Color and dashed contours show the measured $\xi^{\rm Ly\alpha}$, while solid contours show the fitted models. \emph{Panel e:} The projected galaxy-Ly$\alpha$ correlation function $w_p^{\rm Ly\alpha}$. Colored data points correspond to the stellar mass bins indicated in the legend and are horizontally offset for clarity. Colored lines (heavily overlapping) show the corresponding model for each bin.}
    \label{fig:gallya_fourbin_model}
\end{figure*}

\subsection{Results: Mass-Dependent Galaxy-Ly$\alpha$ Clustering}

We now split the foreground galaxies into 4 bins of stellar mass, listed in Table~\ref{tab:gallyamasses}. The bins have the same stellar mass ranges as those used in Section~\ref{sec:galgalclustering}, except that the last dividing point is lower by 0.1 dex. We made this slight adjustment because we found that a usefully precise measurement required the highest-mass bin to be slightly wider, owing to the smaller galaxy sample available for the Ly$\alpha$ analysis.

Measurements of $\xi^{\rm Ly\alpha}$ and $w^{\rm Ly\alpha}_p$ are shown in Fig.~\ref{fig:gallya_fourbin_model} for each stellar mass bin. Visually, we see more extended absorption around galaxies in the two more massive bins (panels c and d) than in the lower-mass bins (panels a and b), which is reflected in $-w_p$ being generally higher (panel e) in the higher-mass bins at large $r_p$. However, the covariance in $\xi^{\rm Ly\alpha}$ at large $r_p$ makes it impossible to gauge the significance of such differences ``by eye.'' 

In fact, we infer statistically compatible halo masses in all stellar mass bins, as listed in Table~\ref{tab:gallyamasses}. We will discuss the significance of this result in Section~\ref{sec:discussion}. The models again produce an acceptable fit quality, as judged by $\chi^2$, in all cases. We found that these mass estimates were not significantly changed by fitting each stellar mass bin independently (thus removing inter-bin covariances in $\xi^{\rm Ly\alpha}$ as well as decoupling $\sigma_z$) or by fitting $\xi^{\rm Ly\alpha}$ over a reduced range of $r_p$ and $\pi$. Turning to the projected $w_p^{\rm Ly\alpha}$, we found again that 1D fits produced noisier but consistent $M_h$ estimates, which are listed in the last column of Table~\ref{tab:gallyamasses}. Corner plots for both fits are shown in Appendix~\ref{sec:corners}. 

\section{Discussion}
\label{sec:discussion}
    
\begin{figure*}
    \centering
    \includegraphics[width=3.5in]{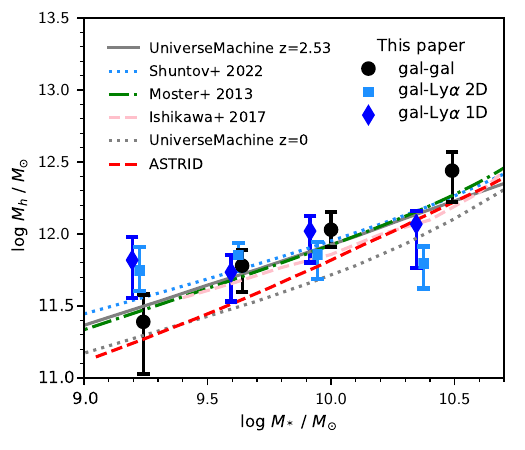}
    \includegraphics[width=3.5in]{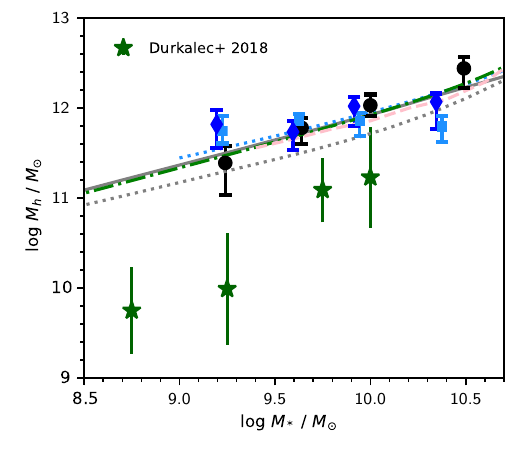}
    \caption{\emph{Left panel:} The stellar mass-halo mass (SMHM) relation at $z=2.5$, derived in this paper using galaxy-galaxy and galaxy-Ly$\alpha$ clustering (symbols indicated in legend), is compared to several literature results. The galaxy-Ly$\alpha$ points are slightly offset in $\log M_*$ for clarity. The \citet{Shuntov22} results are based on $M_*$- and redshift-dependent abundance and clustering measurements using the COSMOS2020 photometric catalog; we interpolate their redshift bins to $z=2.5$. The \citet{Ishikawa17} relation is derived from the $M_*$-dependent angular correlation functions of $u$-band dropouts at $z \sim 3$ in Canada-France-Hawaii Telescope Legacy Survey imaging. The UniverseMachine \citep{Behroozi19} and \citet{Moster13} relations are based on self-consistent multi-epoch modeling of the observed stellar mass function (Moster) along with galaxy clustering and other inputs (Behroozi). We evaluate the analytic \citet{Moster13} SMHM relation at $z=2.5$, and we use the UniverseMachine DR1 tabulation of $\langle \log M_h | M_* \rangle$ at $z=2.53$ and $z=0$. The SMHM relation in the ASTRID simulation is shown, using the median $M_h$ given $M_*$ in the $z=2.5$ snapshot. \emph{Right panel:} The same results are shown on an expanded scale with the \citet{Durkalec18} SMHM relation at $z \sim 3$ derived from galaxy-galaxy clustering in VUDS. We plot the HOD parameter $M_{\rm min}$ to represent the halo mass of their $M_*$-thresholded subsamples.}
    \label{fig:smhm}
\end{figure*}

Our constraints on the SMHM relation at $z=2.5$ from both galaxy-galaxy and galaxy-Ly$\alpha$ clustering are shown in Fig.~\ref{fig:smhm} and compared to earlier work. We begin by considering our galaxy-galaxy clustering results. We find moderate evidence of $M_*$-dependent galaxy clustering at $3.5\sigma$ significance: the slope of the SMHM relation, based on a linear fit to the halo masses derived in Section~\ref{sec:galgalclustering}, is $d \log M_h / d \log M_* = 0.85 \pm 0.24$. Both the slope and normalization agree well with a variety of literature results shown in the left panel of Fig.~\ref{fig:smhm} that are ultimately derived (with the exception of the ASTRID curve) from analyses of imaging surveys, primarily $M_*$- and photometric redshift-dependent measures of galaxy abundances and angular correlation functions. Specifically, the average slope of the UniverseMachine \citep{Behroozi19} relation over the $M_*$ range plotted in Fig.~\ref{fig:smhm} is $d \log M_h / d \log M_* = 0.58$, which is representative of the empirical curves plotted. This slope is consistent with our galaxy-galaxy clustering-based slope at the 1$\sigma$ level. The SMHM relation slope in the ASTRID simulation is steeper ($d \log M_h / d \log M_* = 0.73$), closer to our galaxy-galaxy measure but, as we will see, farther from our galaxy-Ly$\alpha$ result.

Our measurements support an evolution in the SMHM relation, since they agree much more closely with the UniverseMachine relation at $z=2.5$ than at $z=0$. Furthermore, the consistency (Fig.~\ref{fig:smhm}) of the halo masses inferred for LATIS galaxies, which consists of fairly UV-bright ($r < 24.8$) LBGs, and for photometric samples suggests that LBG clustering is representative of the full galaxy population at matched $M_*$.

Compared to earlier measures of galaxy two-point correlation functions ($\xi$ and $w_p$) at similar redshifts based on spectroscopic samples, we find good agreement with bulk estimates of the halos hosting similarly selected LBGs, undifferentiated by stellar mass. Our threshold mass for the LATIS sample as a whole, $\log M_h^{\rm thresh} / {\rm M}_{\odot} = 11.56^{+0.06}_{-0.15}$, is consistent at the $<2\sigma$ level with \citet[][$\log M_h^{\rm thresh} / {\rm M}_{\odot} = 11.7 \pm 0.1$]{Trainor12} and \citet[][$\log M_h^{\rm thresh} / {\rm M}_{\odot} = 11.73 \pm 0.07$ and $11.57 \pm 0.15$ for their VLT+Keck and VLT-only samples, respectively]{Bielby13}. Compared to our initial estimate in \citet{Newman22}, $\log M_h^{\rm thresh} / {\rm M}_{\odot} = 11.8 \pm 0.1$, the improved estimate in this paper is about $2\sigma$ lower.\footnote{The main result of \citet{Newman22}, the low galaxy overdensity $\delta_{\rm gal}$ within the regions of strongest large-scale Ly$\alpha$ absorption, is not very sensitive to changes to $M_h^{\rm thresh}$. A forthcoming paper (Newman et al., in prep.) will revisit and expand up on the \citet{Newman22} analysis using the full LATIS maps.}

Prior to this work, the only derivation of the $z \sim 3$ SMHM relation from a spectroscopic survey was performed by \citet{Durkalec15,Durkalec18} based on the VIMOS Ultra-Deep Survey (VUDS). \citet{Durkalec18} found a steep SMHM that would imply an extremely high star-formation efficiency in low-mass, high-redshift galaxies. Although our sample does not reach their lowest-mass bin, it is clear that our galaxy-galaxy and galaxy-Ly$\alpha$ results are not consistent with \cite{Durkalec18} in the three overlapping mass bins.\footnote{The significance of the difference in $\log M_h$ ranges from 1.4-2.0$\sigma$ per $M_*$ bin, with the Durkalec $M_h$ always being lower. The average difference in $\log M_h$ is non-zero at a significance of $\sim$3$\sigma$.} Small differences in the sample selection (e.g., limiting magnitude, mean redshift) are not a plausible explanation for this difference. Although we cannot determine the origin of the difference, we consider that the consistency of our SMHM measures with each other and with the other results in Fig.~\ref{fig:smhm} (left panel) strongly support a more conventional SMHM relation.

Now we turn to the galaxy-Ly$\alpha$ cross-correlation. We modeled and fit both the full 2D $\xi^{\rm Ly\alpha}$ and a projected 1D statistic $w_p^{\rm Ly\alpha}$. Although the velocity structure of the Ly$\alpha$ absorption provides no constraining power on the halo mass, the 2D fits nonetheless give more precise constraints on $M_h$, because they naturally weight the observed absorption as a function of $\pi$. The 1D fits, on the other hand, are insensitive to our modeling of both redshift measurement errors and the velocity structure of the gas, and we further removed sightlines within $r_p < 1$ \cMpch~that are expected to be most sensitive to feedback. Fig.~\ref{fig:smhm} shows that the masses derived from the 2D and 1D analyses are consistent. We do not detect a significant $M_*$-dependence of the surrounding Ly$\alpha$ absorption, but we find that this is only moderately surprising given the uncertainties. Specifically we find slopes of $d \log M_h / d \log M_* = 0.00 \pm 0.16$ and $0.22 \pm 0.22$ for the 2D and 1D fits, respectively. Compared to the UniverseMachine SMHM relation, which is representative of the other models plotted in Fig.~\ref{fig:smhm}, the measured slopes are shallower with significances of $3.5\sigma$ and $1.6\sigma$. Thus the SMHM slope from the 1D fits is consistent with standard models, whereas the 2D fits give a notably shallow slope. The largest source of this flat slope is moderate tension  in the highest-$M_*$ bin between the halo masses inferred from 2D galaxy-Ly$\alpha$ clustering and galaxy-galaxy clustering (as well as the plotted models). This tension is much reduced in the 1D analysis, which could indicate that the velocity structure of the H I around massive galaxies in not correct in the simulations. However, the size of the uncertainties prevents any firm conclusion.

Compared to earlier studies of the galaxy-Ly$\alpha$ cross-correlation, our results are considerably more precise and are the first to quantitatively estimate halo masses in multiple $M_*$ bins. For instance, our $1\sigma$ uncertainty in $\log M_h^{\rm thresh}$ is 0.09~dex compared to 0.2~dex and 0.6~dex in the \citet{Rakic13} and \citet{Kim08} studies, respectively. This is a statistical error, which does not incorporate uncertainties in the estimation of the halo-Ly$\alpha$ cross-correlation from simulations. \citet{Rakic13} found that the inclusion of AGN feedback in the OWLS simulations shifted their inferred halo mass by 0.3~dex. They regarded this as a mild uncertainty since it was comparable to their statistical errors, but it would be substantially larger than ours. However, we expect the effect of AGN feedback on our halo-Ly$\alpha$ cross-correlation to be much smaller. First, our measurements focus on larger scales; second, the AGN feedback implementation in OWLS was very strong. This can be seen, for instance, by comparing its dramatic effect on the star-formation rate density \citep[][Fig.~18]{Schaye10} to the subtle effect in the more recent simulations ASTRID, TNG, and SIMBA \citep[][Fig.~4]{Ni23}.

We clearly detect redshift-space distortions in the 2D $\xi^{\rm Ly\alpha}$ maps, with substantial compression in the $\pi$ direction that we attribute to large-scale gas infall onto overdensities. Such infall has been previously observed around similar galaxy samples \citep{Rudie12,Rakic13,Tummuangpak14,Bielby17,Chen20}. We found that the detailed shape of $\xi^{\rm Ly\alpha}$, especially the pinching at small $r_p$, is influenced by our spectral processing steps, specifically MFR. By carefully matching these steps in our synthesis of halo-Ly$\alpha$ cross-correlations models, we match the observed shape closely. Although the observed redshift-space distortions are not diagnostic of the halo mass and so are not a focus of this paper, they support the precision of the LATIS redshifts ($\sigma_z = 66 \pm 8$~\kms~in our $\xi^{\rm Ly\alpha}$ fits).

Altogether, we find a consistent picture in which our $M_*$-dependent galaxy-galaxy and galaxy-Ly$\alpha$ correlation function measurements are accurately modeled by careful treament of cosmological simulations, leading to estimates of the SMHM that are generally consistent with each other and with canonical models largely based on photometric surveys. This is significant for several reasons. First, future analyses of the LATIS maps will rely on understanding the halo population occupied by our galaxy sample. A prime example is \citet{Newman22}, in which observed galaxy overdensities in different large-scale environments, as traced using 3D Ly$\alpha$ absorption maps, were compared to halo overdensities at matched $M_h$ and environment in simulations.

Second, comparing halo masses estimated from galaxy-galaxy and galaxy-Ly$\alpha$ clustering provides a strong test of our ability to measure and model subtle differences in Ly$\alpha$ absorption. The most sensitive comparison is between the two analyses of the full LATIS galaxy sample. Galaxy-galaxy clustering led to a threshold mass of $\log M_h / {\rm M}_{\odot} = 11.56^{+0.06}_{-0.15}$, and galaxy-Ly$\alpha$ clustering (2D) gave $11.54 \pm 0.09$. However, as discussed in Section~\ref{sec:hmdef}, we expect small differences due to the different cosmological parameters and subhalo finders employed by the two simulations underlying these analyses. Based on Fig.~\ref{fig:hmdef}, the mass-thresholded halo population with $\log M_h^{\rm thresh} / {\rm M}_{\odot} = 11.54$ in ASTRID has the same autocorrelation function the halo population with $\log M_h^{\rm thresh} / {\rm M}_{\odot} > 11.47$ in MDPL2. Correcting for this known difference, we find that the threshold masses differ by $\Delta \log M_h = 0.09^{+0.11}_{-0.17}$. Here we have estimated the uncertainty assuming that errors in the galaxy-galaxy and galaxy-Ly$\alpha$ results are uncorrelated; the true error in $\Delta \log M_h$ must be smaller. Thus we find consistency at about the 0.1 dex level. This corresponds to 4\% multiplicative changes in $\xi^{\rm Ly\alpha}$ and the $\delta_F$ measurements that underlie it, suggesting that we are able to measure and model Ly$\alpha$ absorption to a fairly high precision. Note for typical $w_p^{\rm Ly\alpha} \approx -0.1$ at $r_p \approx 4$~\cMpch, 4\% multiplicative errors correspond to $\Delta \delta_F = 0.004$.

Third, we find that galaxy-Ly$\alpha$ clustering gives similarly precise constraints on halo occupation as galaxy-galaxy clustering. We initially suspected that this conclusion would be somewhat particular to LATIS, since it is designed around deep exposures required to measure the Ly$\alpha$ forest in galaxy spectra. However, we found that increasing the random errors in $\delta_F$ by a factor of $\sqrt{2}$ when constructing our covariance matrix, effectively halving the exposure time, would increase the uncertainty in $\log M_h$ by only 10\%. This modest increase indicates that intrinsic Ly$\alpha$ correlations are the major part of the error budget, and a wider, shallower survey would be more powerful for global correlation analyses. This suggests that galaxy-Ly$\alpha$ correlations come ``for free'' and will be a powerful application of any future large optical spectroscopic surveys of high-redshift galaxies.

\section{Summary}
\label{sec:summary}

We constrained the galaxy-halo connection at $z = 2.5$ through measurements of both galaxy two-point correlation functions and the cross-correlation between galaxies and transverse Ly$\alpha$ absorption, as detected in background galaxy spectra. The measurements are based on $\sim$3000 spectra collected by the LATIS survey. We presented a new method of measuring all auto- and cross-correlations among galaxies binned according to stellar mass, and by analyzing mock surveys within the MultiDark MDPL2 simulation, we derived a covariance matrix and estimated the typical halo mass associated with each bin. We then measured the galaxy-Ly$\alpha$ cross-correlation, both for the galaxy sample as a whole and in bins of stellar mass. We generated a large sample of synthetic Ly$\alpha$ spectra in the ASTRID simulation, carefully matched to the characteristics of LATIS spectra, and inferred halo masses by comparing the simulated halo-Ly$\alpha$ and observed galaxy-Ly$\alpha$ cross-correlations, both in two and one dimensions.

We found that the constraints from the two methods are consistent with one another and with standard SMHM relations derived from photometric surveys. We did not find evidence of an unusually steep SMHM relation \citep{Durkalec18} or of differences between the simulated and observed galaxy-Ly$\alpha$ cross-correlation functions \citep{Momose21b}. These results will inform future work on the galaxy-IGM connection using the LATIS survey data, and they tightly constrain systematic errors in our measurements and modeling of the Ly$\alpha$ transmission fluctuations. Our results also highlight galaxy-Ly$\alpha$ clustering as a tool whose power will increase with future large optical spectroscopic surveys of the distant universe. 

\begin{acknowledgments}
This paper includes data gathered with the 6.5-meter Magellan Telescopes located at Las Campanas Observatory, Chile. We gratefully acknowledge the support of the Observatory staff. We thank the anonymous referee for thoughtful comments that improved the manuscript. This material is based upon work supported by the National Science Foundation under Grant No.~AST-2108014. B.C.L. acknowledges the support from the National Science Foundation under Grant No.~1908422. S.B.~was supported by NASA-80NSSC21K1840. M.Q.~is supported by NSF grant AST-2107821. The authors acknowledge the Frontera computing project at the Texas Advanced Computing Center (TACC) for providing HPC and storage resources that have contributed to the research results reported within this paper. Frontera is made possible by NSF OAC-1818253.
\end{acknowledgments}

\clearpage

\bibliography{main}{}
\bibliographystyle{aasjournal}


\begin{appendix}

\section{Corner Plots}
\label{sec:corners}

\begin{figure*}
    \centering
    \includegraphics[width=0.7\linewidth]{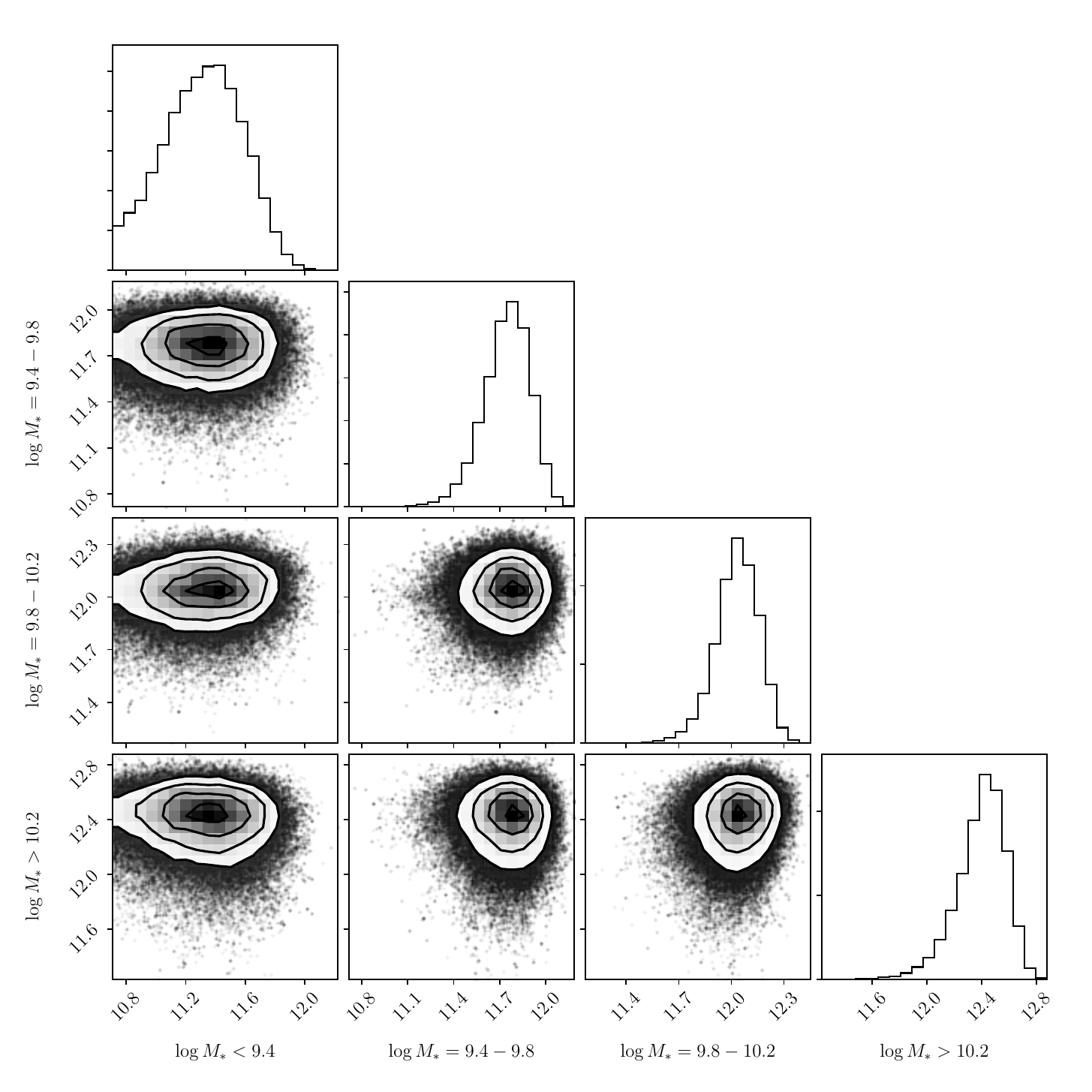}
    
    \caption{Corner plot illustrating posteriors for our model fit to the stellar mass-dependent galaxy-galaxy clustering data in Fig.~\ref{fig:crosscorr}. The four parameters correspond to the halo mass $\log M_h / {\rm M}_{\odot}$ for LATIS galaxies in the stellar mass intervals indicated. Contours show the 1, 2, and 3$\sigma$ levels while fully marginalized posteriors are shown at the top of each row.}
    \label{fig:crosscorrcorner}
\end{figure*}

\begin{figure*}
    \centering
    \includegraphics[width=0.4\linewidth]{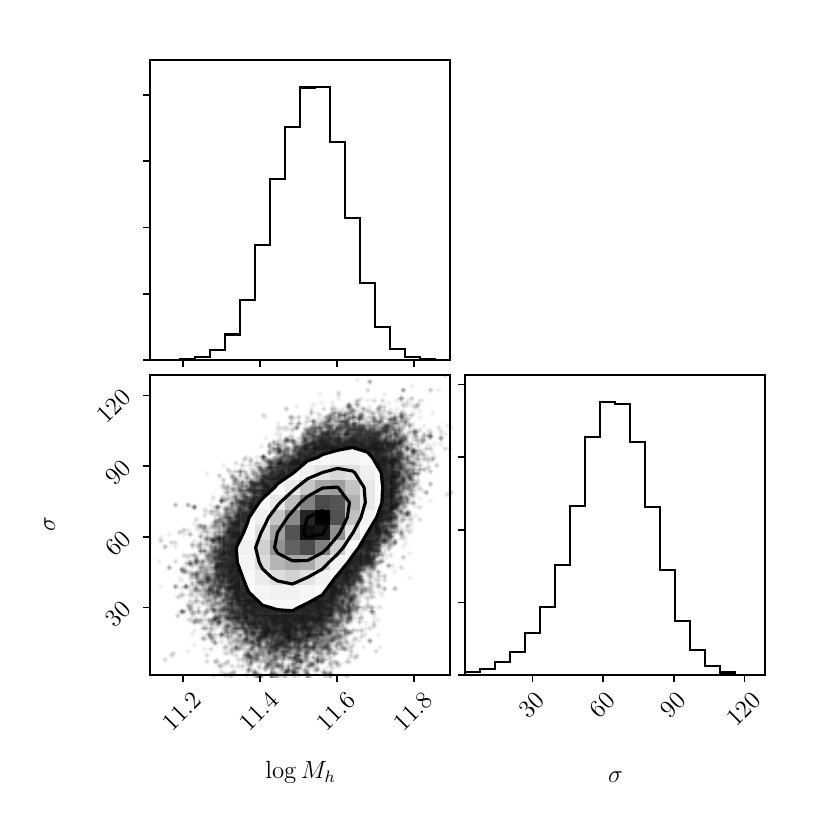}
    \caption{Corner plot illustrating posteriors for our fit to the 2D galaxy-Ly$\alpha$ cross-correlation $\xi^{\rm Ly\alpha}$ for the entire LATIS sample; data are shown in Fig.~\ref{fig:gallya_onebin_model}.}
    \label{fig:gallya_onebin_corner}
\end{figure*}

\begin{figure*}
    \centering
    \includegraphics[width=0.7\linewidth]{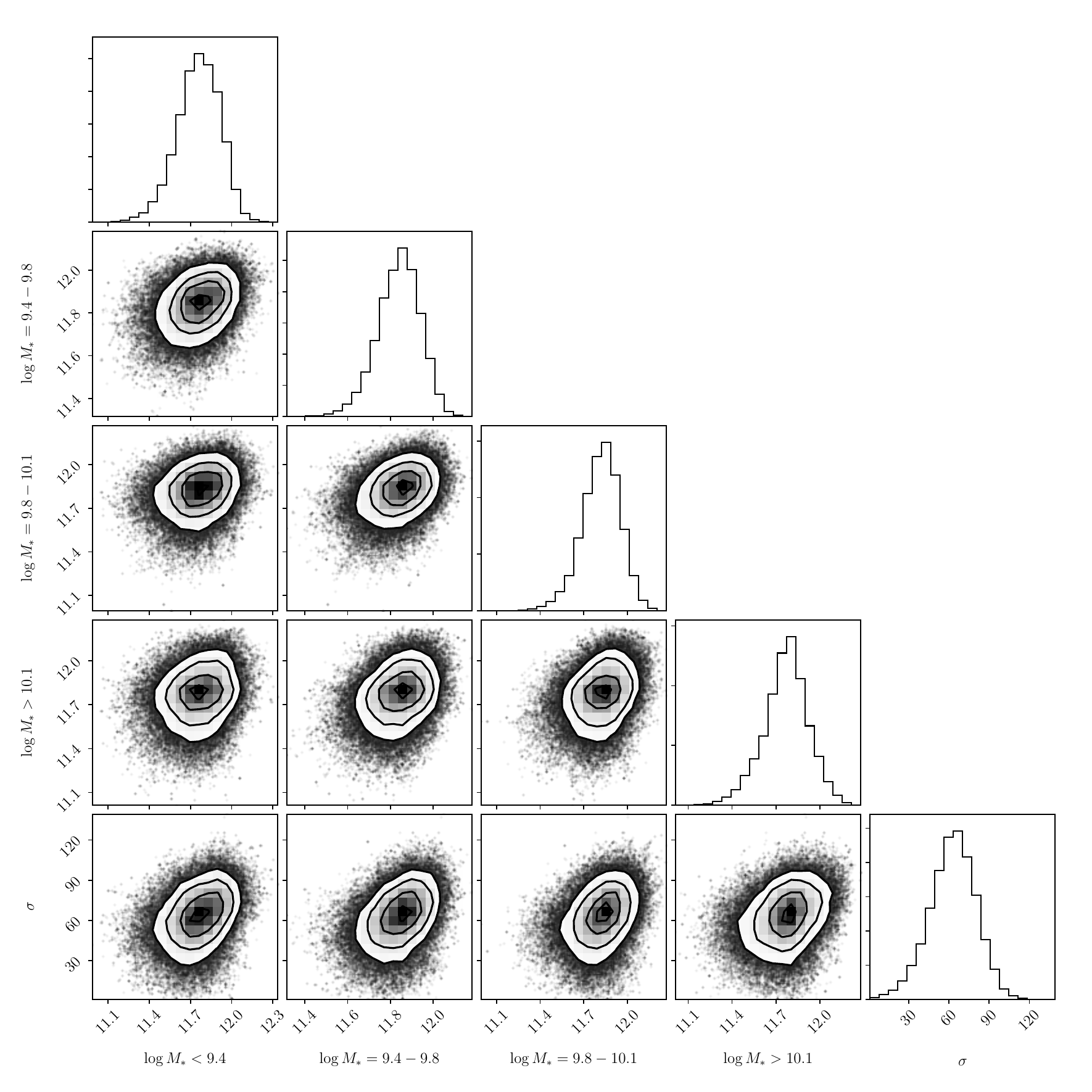}
    \caption{Corner plot illustrating posteriors for our fit to the 2D galaxy-Ly$\alpha$ cross-correlation $\xi^{\rm Ly\alpha}$ in four bins of stellar mass. Similar to Fig.~\ref{fig:crosscorrcorner}, each parameter describes the halo mass $\log M_h / {\rm M}_{\odot}$ for LATIS galaxies in the stellar mass intervals indicated.}
    \label{fig:gallya_fourbins_corner}
\end{figure*}

Figures \ref{fig:crosscorrcorner}-\ref{fig:gallya_fourbins_corner} demonstrate one- and two-dimensional marginalized posteriors for our models fit to galaxy-galaxy and galaxy-Ly$\alpha$ clustering data. 

\clearpage

\section{Angular Variation of Target Sampling and Spectroscopic Success Rates}
\label{sec:esr}

\begin{figure}
    \centering
    \includegraphics[width=\linewidth]{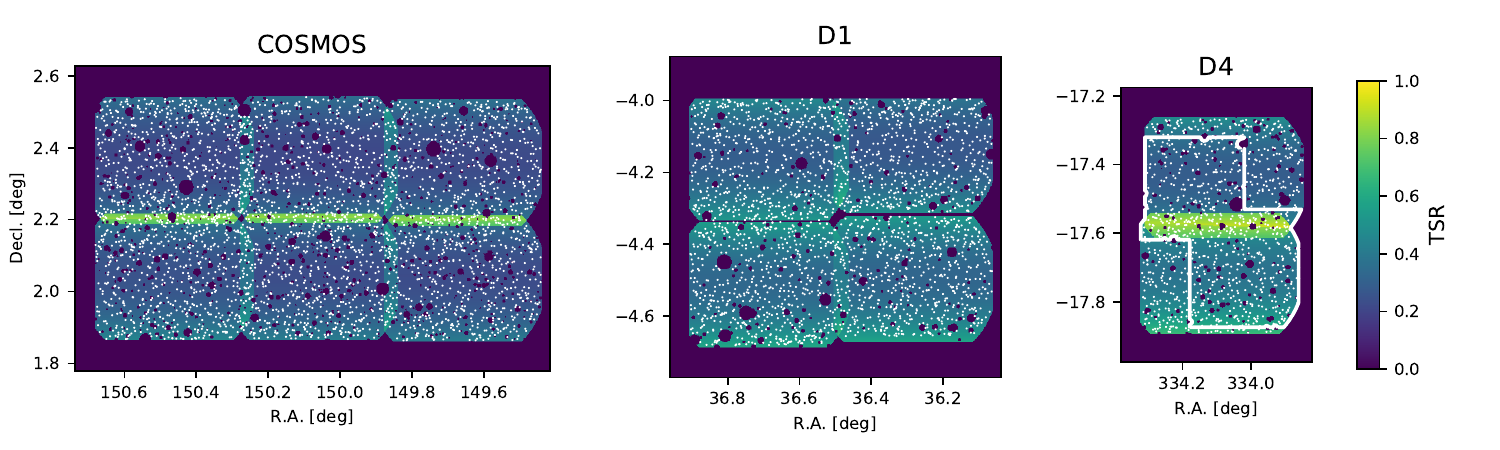}
    \includegraphics[width=\linewidth]{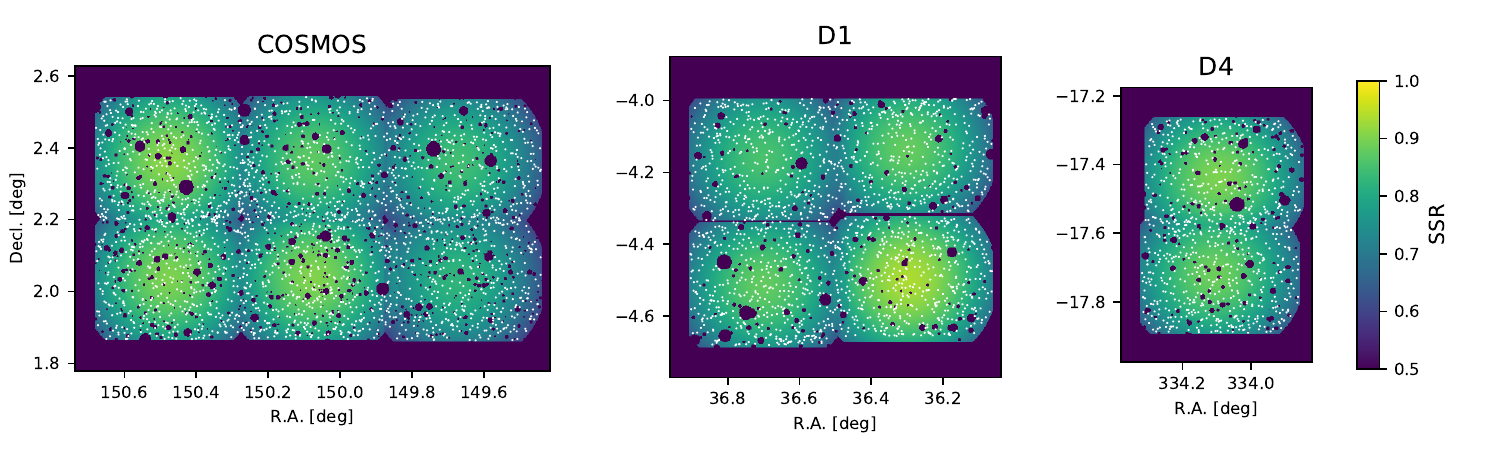}
    \caption{\emph{Top row:} The variation of the target sampling rate (TSR) within each of the LATIS survey fields is indicated by the color bar. White points show positions of targeted galaxies. The white contour in the D4 field shows the subregion covered by near-infrared imaging that is used in this paper. \emph{Bottom row:} Variation of the spectroscopic sampling rate (SSR), with white points showing the position of galaxies with secure redshifts. Note the compressed color scale.}
    \label{fig:surveymasks}
\end{figure}

The target sampling rate (TSR) and spectroscopic success rate (SSR) vary within the LATIS survey footprint. Here we describe a model of the angular dependence of these quantities.

\subsection{Target Sampling Rate}

We first consider the TSR. We built an initial model of the area surveyed by each of the 12 LATIS footprints, based on the IMACS field of view excluding masked regions surrounding bright stars. Note that the FOV has an asymmetric shape in the east-west direction due to a lateral shift of the dispered spectra as described by \citet{Newman20}. We note that in the D1 field, the CFHTLS imaging boundary sets the northern edge. We created a model ${\rm TSR}_{\rm foot}$ that accounts for variation of the TSR from footprint to footprint and in regions of overlapping footprints. 
First, within the unmasked and non-overlapping portion of each footprint, we compute the TSR and assign this to ${\rm TSR}_{\rm foot}$. For this calculation, and throughout this paper, we exclude galaxies that were observed only as part of a ``bright target'' mask (see \citealt{Newman20}) intended for poor weather conditions, since these masks do not cover the full survey area and were not uniformly observed. Second, within each field, we calculated the average TSR within regions where two footprints overlap. We considered separately overlaps in the E-W and N-S directions, since we attempted to reobserve objects in the E-W overlap regions more frequently to partly compensate for the poorer image quality at the edge of the IMACS field (see below). We found that the individual E-W or N-S overlap regions within each field were consistent with having the same TSR, and therefore set ${\rm TSR}_{\rm foot}$ to the mean to reduce noise.

\begin{figure}
    \centering
    \includegraphics[width=0.4\linewidth]{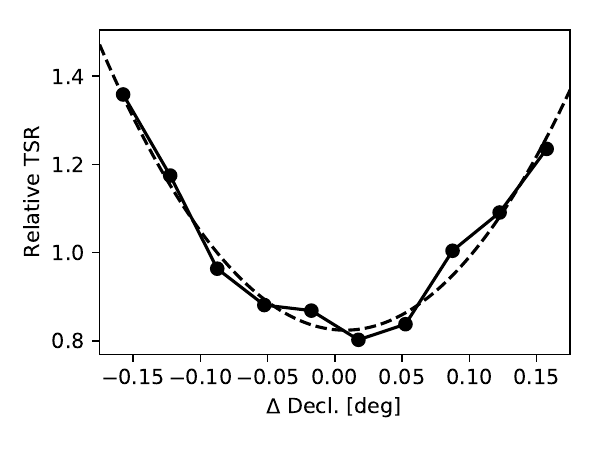}
    \includegraphics[width=0.4\linewidth]{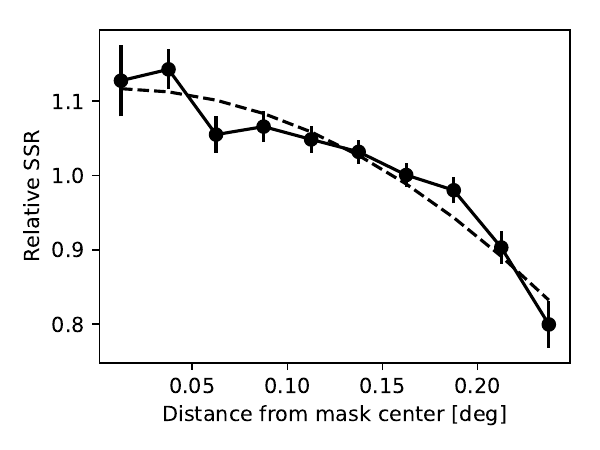}
    \caption{\emph{Left:} Variation of ${\rm TSR}_{\rm mask}$ with the declination of a target relative to the mask center. Black points and solid line show the measurements, averaged over all footprints, while the dashed line shows the quadratic fit used in our modeling. \emph{Right:} Variation of ${\rm SSR}_{\rm mask}$ with radial distance of a target relative to the mask center, along with a quadratic fit as in the left panel.}
    \label{fig:residvar}
\end{figure}

We then considered the probability of target selection as a function of position relative to the footprint center, ${\rm TSR}_{\rm mask}$. In computing this refined TSR, we weighted targets by the inverse of ${\rm TSR}_{\rm foot}$ in order to quantify the residual TSR variation. We found that ${\rm TSR}_{\rm mask}$ is primarily a function of vertical (i.e., N-S) position relative to the mask center, as shown in Fig.~\ref{fig:residvar}, such that the TSR is higher near the top and bottom of a footprint than at the midline. This dependence arises from the direction of dispersion (N-S) and the requirement that spectra not overlap, which tends to sort targets into ranks. We modeled ${\rm TSR}_{\rm mask}$ as a quadratic function of vertical mask position. The final TSR model is then the product of ${\rm TSR}_{\rm foot}$ and ${\rm TSR}_{\rm mask}$.

Figure~\ref{fig:surveymasks} shows our model of the TSR within each field. The median TSR is lower in COSMOS (0.28) than in D1 (0.37) and D4 (0.39) because the parent sample is larger: we include both $z_{\rm phot}$- and $ugr$-selected galaxies in COSMOS and only the latter in D1 and D4. For the purpose of this paper, only the relative variation of the TSR within each field is relevant, but for other applications, it may be important to recognize that the TSR depends on magnitude and is significantly higher for galaxies with $r < 24.4$ (see \citealt{Newman20}). The NMAD (normalized median absolute deviation) of the TSR within each field is 6-10\% of the median, so despite the structure seen in the Figure, the overall target sampling is fairly uniform within LATIS fields.

In this paper, an additional criterion for a galaxy to be included in our sample is the availability of adequately deep near-infrared imaging. Such imaging in the D4 field is shallower than in the other fields and does not cover the entire LATIS area. We found that galaxies outside the white contour in Fig.~\ref{fig:surveymasks} lack a NIR detection much more frequently than in the rest of the D4 field, and we therefore do not consider the region outside the contour in this paper.

\subsection{Spectroscopic Success Rate}

We now consider the SSR, defined as the fraction of targeted sources for which a high-confidence redshift was determined ({\tt zqual} = 3 or 4). Low-confidence redshifts are not used in this paper. The most significant spatial dependence of the SSR was found to be the distance of a target from the footprint center. This dependence arises because the size of images produced by the IMACS f/2 camera increases significantly with field radius, which reduces the signal-to-noise ratio of the spectra. The right panel of Figure~\ref{fig:residvar} shows the relative variation of the SSR with this radial distance, defined to average to unity. We use a quadratic fit to estimate ${\rm SSR}_{\rm mask}$. Galaxies with increased exposure due to repeat observations were not included in this calculation in order to isolate the radial dependence of SSR from any spatial dependence of exposure time, which will be considered in the next step.

We then consider large-scale variations ${\rm SSR}_{\rm foot}$, which may arise from different observing conditions when different footprints are observed, or from the fact that some sources that fall within two footprints receive a longer integration time. When calculating ${\rm SSR}_{\rm foot}$, we weight galaxies by the inverse of ${\rm SSR}_{\rm mask}$ in order to determine the residual spatial variations. We calculate the weighted SSR within each footprint, first considering only the non-overlapping part, and find that footprint-to-footprint variations in LATIS are small (mean 0.79, standard deviation 0.026). We next compared the weighted SSR within overlapping and non-overlapping regions, and found that SSR differences were small (a few percent) and within the expected Poisson fluctuations. Since the overlap regions are small and the overall variation in SSR is weak, we simply set ${\rm SSR}_{\rm foot}$ within overlap regions to the average value of the two footprints.

The overall SSR is given by the product of ${\rm SSR}_{\rm mask}$ and ${\rm SSR}_{\rm foot}$ and is shown in Figure~\ref{fig:surveymasks}. The effective sampling rate (ESR) is the product of the TSR and SSR.

\end{appendix}

\end{document}